\documentclass[preprint,amsmath,amssymb,aps,prc]{revtex4-2}

\usepackage{graphicx}
\usepackage{dcolumn}
\usepackage{bm}
\usepackage{epstopdf}
\usepackage[mathlines]{lineno}
\usepackage{harpoon}
\usepackage{makecell}
\usepackage{subfigure}
\usepackage{hyperref}
\usepackage{color}
\usepackage{braket}
\usepackage{amsmath}
\usepackage{mathrsfs}
\usepackage{appendix}
\usepackage{multirow}
\usepackage{float}
\usepackage{orcidlink}

\begin{document}

\title{Triaxial relativistic Hartree-Bogoliubov theory in continuum for exotic nuclei}

\author{{K. Y. Zhang}\orcidlink{0000-0002-8404-2528}}
\affiliation{National Key Laboratory of Neutron Science and Technology, Institute of Nuclear Physics and Chemistry, China Academy of Engineering Physics, Mianyang, Sichuan 621900, China}
\affiliation{State Key Laboratory of Nuclear Physics and Technology, School of Physics, Peking University, Beijing 100871, China}

\author{{S. Q. Zhang}\orcidlink{0000-0002-9590-1818}}
\affiliation{State Key Laboratory of Nuclear Physics and Technology, School of Physics, Peking University, Beijing 100871, China}

\author{{J. Meng}\orcidlink{0000-0002-0977-5318}} \email{mengj@pku.edu.cn}
\affiliation{State Key Laboratory of Nuclear Physics and Technology, School of Physics, Peking University, Beijing 100871, China}

\date{\today}

\begin{abstract}
  A triaxial relativistic Hartree-Bogoliubov theory in continuum (TRHBc) has been developed to incorporate triaxial deformation, pairing correlations, and continuum effects in a fully microscopic and self-consistent way, aiming for a reliable description of triaxial exotic nuclei with extreme neutron-to-proton ratios.
  The TRHBc formalism is presented in detail, and its numerical implementation is benchmarked against the results from the axially deformed relativistic Hartree-Bogoliubov theory in continuum and the TRHB theory in harmonic oscillator expansion.
  The TRHBc theory is applied to investigate the aluminum isotopes systematically, and the available data are well reproduced for the binding energies, one- and two-neutron separation energies, and charge radii.
  The nuclei near the one-neutron drip line, $^{40}$Al and $^{42}$Al, are found to be triaxially deformed with one-neutron separation energies below 1 MeV.
  Possible neutron halos in the triaxial nuclei $^{40}$Al and $^{42}$Al are explored by examining the single-particle levels around the Fermi surface, including their composition and contribution to the total neutron density.
  The existence of neutron halos in $^{40}$Al and $^{42}$Al is also supported by the halo scale, which is comparable to other halo nuclei well-established previously.
  More importantly, the mechanism for the halo formation in $^{40}$Al is revealed to be the triaxial deformation, which results in the decoupling of the halo orbitals from those of the core.
\end{abstract}

\date{\today}

\maketitle

\section{Introduction}

The triaxial relativistic Hartree-Bogoliubov theory in continuum (TRHBc) is a density functional theory formulated covariantly~\cite{Meng2016Book}, which can take into account simultaneously triaxiality, pairing correlations, and continuum effects in a fully microscopic and self-consistent way.
In Ref.~\cite{Zhang2023PRC(L2)}, the TRHBc theory is briefly reported and applied to predict $^{42}$Al as a triaxial halo nucleus.
Here in this paper, the TRHBc formalism and numerical details will be presented in detail, the aluminum isotopes will be investigated systematically, and the possible emergence of halo nuclei driven by triaxial deformation will be particularly emphasized.

The observation of an anomalous enhancement in the cross section for reactions of $^{11}$Li with light target nuclei in 1985~\cite{Tanihata1985PRL}, later identified as evidence of a nuclear halo phenomenon~\cite{Hansen1987EPL,Kobayashi1988PRL}, has opened a new era for nuclear physics.
This discovery stimulated the construction of next-generation rare isotope beam facilities and the observation of other novel phenomena in exotic nuclei including the change of magic numbers~\cite{Ozawa2000PRL} and the pygmy resonance~\cite{Adrich2005PRL}.
So far, about 20 halo nuclei or candidates have been identified experimentally~\cite{Tanihata2013PPNP}.
Theoretically, lots of efforts have been made to describe and predict halo nuclei based on the few-body models~\cite{Zhukov1993PhysRep,Hansen1995ARNPS}, the shell model~\cite{Otsuka1993PRL,Kuo1997PRL}, antisymmetrized molecular dynamics~\cite{Horiuchi1994ZPA,Itagaki1999PRC}, halo effective field theory~\cite{Ryberg2014PRC,Ji2014PRC}, \textit{ab initio} methods~\cite{Calci2016PRL,Shen2025PRL}, and nonrelativistic~\cite{Terasaki1996NPA,Schunck2008PRC} and relativistic~\cite{Meng1996PRL,Meng1998PRL,Meng2015JPG} density functional theories.

The relativistic or covariant density functional theory (CDFT) has proven to be a successful framework for describing both nuclear ground-state and excited-state properties~\cite{Ring1996PPNP,Vretenar2005PhysRep,Meng2006PPNP,Niksic2011PPNP,Meng2013FOP,Meng2015JPG,Meng2016Book,Zhou2016PhysScr,Shen2019PPNP,Zhang2024NPR}.
The CDFT offers numerous appealing advantages~\cite{Ring2012PhysScr,Meng2021AAPPS}, such as the automatic inclusion of the spin-orbit interaction~\cite{Koepf1991ZPA,Ren2020PRC(R)} and the reasonable description of its isospin dependence~\cite{Sharma1995PRL}, the new saturation mechanism arising from the competition between scalar and vector densities~\cite{Walecka1974AP}, the natural explanation of the pseudospin symmetry in the nucleon spectrum~\cite{Ginocchio1997PRL,Meng1998PRC(R),Liang2015PhysRep} and the spin symmetry in the antinucleon spectrum~\cite{Liang2015PhysRep,Zhou2003PRL,He2006EPJA}, and the self-consistent treatment of nuclear magnetism~\cite{Koepf1989NPA,Koepf1990NPA}.
As a result, the CDFT has garnered considerable attention and stands as one of the most widely used nuclear theories.

Based on the CDFT, the relativistic continuum Hartree-Bogoliubov (RCHB) theory was developed, which includes pairing correlations and continuum effects and provides a microscopic and self-consistent description for the neutron halo in $^{11}$Li~\cite{Meng1996PRL}.
In contrast, previous studies within nonrelativistic and relativistic mean-field frameworks could only reproduce the halo features of $^{11}$Li by artificially modifying the neutron orbital~\cite{Bertsch1989PRC,Sagawa1992PLB,Zhu1994PLB}.
This highlights the essential role of pairing correlations and continuum couplings in the theoretical description of halo nuclei.
The RCHB theory has been extended to describe medium-heavy and heavy nuclei, thereby predicting giant halos composed of more than two neutrons~\cite{Meng1998PRL,Meng2002PRC(R),Zhang2002CPL,Zhang2003SciChina,Zhou2024Particles} and new magic numbers in superheavy nuclei~\cite{Zhang2005NPA}.
Based on nuclear densities from the RCHB theory, the interaction cross sections of sodium isotopes and the charge-changing cross sections of carbon, nitrogen, oxygen, and fluorine isotopes on a carbon target have been reproduced by the Glauber model~\cite{Meng1998PLB,Meng2002PLB}, and the elastic proton-nucleus scattering cross sections from light to heavy nuclei have been reproduced in the relativistic impulse approximation~\cite{Kuang2023EPJA}.
The first relativistic nuclear mass table including continuum effects has been constructed using the RCHB theory~\cite{Xia2018ADNDT}.
Based on the RCHB mass table, the $\alpha$-decay energies have been systematically investigated~\cite{Zhang2016CPC}, and the proton radioactivity has been studied for nuclei with $53\le Z\le 83$~\cite{Lim2016PRC}.
Recently, the RCHB mass table has been combined with the kernel ridge regression approach to examine machine learning for assessing physical effects~\cite{Du2023CPC} construct a high-precision nuclear mass model~\cite{Wu2024PRC,Guo2024PRC}.

The existence of halos in deformed nuclei has remained under debate for decades~\cite{Tanihata1995NPA,Misu1997NPA,Hamamoto2004PRC(R),Nunes2005NPA,Zhou2010PRC(R)}.
For instance, all nuclei at the drip line were suggested to be spherical based on a spherical Woods-Saxon (WS) potential~\cite{Tanihata1995NPA}, and a three-body calculation indicated that the halo formation in deformed drip-line nuclei is unlikely~\cite{Nunes2005NPA}.
Given the success of the RCHB theory, it is highly demanded to incorporate deformation, pairing correlations, and continuum effects self-consistently within a unified framework.
This has motivated the development of the deformed relativistic Hartree-Bogoliubov theory in continuum (DRHBc)~\cite{Zhou2010PRC(R),Li2012PRC,Chen2012PRC,Li2012CPL}.
Although $^{40}$Mg is the heaviest magnesium isotope observed so far~\cite{Baumann2007Nature}, the DRHBc theory has predicted the halo phenomena in deformed nuclei $^{42}$Mg and $^{44}$Mg with shape decoupling between the core and the halo~\cite{Zhou2010PRC(R),Li2012PRC}.
The DRHBc theory has also predicted the stability peninsula beyond the primary neutron drip line~\cite{Zhang2021PRC(L),Pan2021PRC,He2021CPC,He2024PRC}, examined the possibility of a proton halo in $^{22}$Al~\cite{Zhang2024PRC,Panagiota2025arXiv}, described successfully the halo nuclei $^{17,19}$B~\cite{Yang2021PRL,Sun2021PRC(1)}, $^{15,19,22}$C~\cite{Sun2018PLB,Sun2020NPA,Wang2024EPJA}, $^{12}$N~\cite{Zhang2025}, $^{31}$Ne~\cite{Zhong2022SciChina,Pan2024PLB}, and $^{37}$Mg~\cite{Zhang2023PLB,An2024PLB}, and predicted deformed neutron halos in $^{39,41}$Na~\cite{Zhang2023PRC(L1)}.
Furthermore, the DRHBc theory has been applied to investigate the connection between halo phenomena and nucleons in the classically forbidden region~\cite{Zhang2019PRC}, the deformation effects on the neutron drip line~\cite{In2021IJMPE}, the shape coexistence~\cite{In2020JKPS,Choi2022PRC,Kim2022PRC,Mun2023PLB,Mun2024PRC.110.024310,Mun2025Particles32} and prolate-shape dominance~\cite{Guo2023PRC}, the shell evolution~\cite{Zheng2024CPC,Zhang2024PhysRevC.110.024302,Wang2024Particles,Pan2025Particles,Huang2025PRC}, the $\alpha$-decay half-lives~\cite{Choi2024PRC}, the nuclear charge radii~\cite{Pan2025PRC}, the inner fission barriers~\cite{Zhang2024CPC}, and the $r$ process~\cite{Pan2025arXiv_mass}.
The DRHBc mass table including both deformation and continuum effects is in progress~\cite{Zhang2020PRC,Zhang2021CSB,Zhang2022ADNDT,Pan2022PRC,Guo2024ADNDT,Zhang2025AAPPS}.
Based on the DRHBc theory, the one-proton emissions in $^{148\text{--}151}$Lu have been studied by the Wentzel-Kramers-Brillouin approximation~\cite{Xiao2023PLB,Lu2024PLB}, the charge-changing cross sections of several $p$-shell nuclei on a carbon target have been reproduced by the Glauber model~\cite{Zhao2023PLB,Zhao2024PLB}, and the beyond-mean-field effects have been explored by the two-dimensional collective Hamiltonian~\cite{Sun2022CPC,Zhang2023PRC}, the angular momentum projection~\cite{Sun2021SciBull,Sun2021PRC(2),Sun2024NPR}, and the finite amplitude method~\cite{Sun2022PRC}.

The nuclear shape deviating from axial symmetry, i.e., the triaxial deformation, may lead to interesting nuclear phenomena such as wobbling motion~\cite{Bohr1975Book}, chiral rotation~\cite{Frauendorf1997NPA}, and the reduction of fission barrier~\cite{Pashkevich1969NPA,Girod1983PRC,Abusara2010PRC,Lu2012PRC(R)}.
It is therefore natural and compelling to explore the possibility of halo structures in triaxially deformed nuclei~\cite{Uzawa2021PRC(L),Zhang2023PRC(L2),Wang2025PRC}.

Incorporating self-consistently triaxial deformation, pairing correlations, and continuum effects is not trivial.
In the RCHB theory~\cite{Meng1998NPA}, the spherical relativistic Hartree-Bogoliubov (RHB) equations are solved in coordinate space.
In the DRHBc theory~\cite{Zhou2010PRC(R)}, the deformed RHB equations are solved in a Dirac WS (DWS) basis~\cite{Zhou2003PRC}, and axial deformation is taken into account by expanding nuclear densities in Legendre polynomials.
To further include triaxial deformation, while the triaxial RHB equations can still be solved in the DWS basis, the nuclear densities should be expanded in terms of the more sophisticated spherical harmonics.
The triaxial relativistic Hartree-Bogoliubov theory in continuum thus developed has been briefly reported in Ref.~\cite{Zhang2023PRC(L2)}.

In this paper, the TRHBc formalism is presented in detail in Sec.~\ref{theory}, and its numerical implementation is benchmarked against the results from the DRHBc theory and the TRHB theory in harmonic oscillator expansion in Sec.~\ref{numerical}.
The TRHBc theory is applied to investigate the aluminum isotopes systematically in Sec.~\ref{results}.
A summary is given in Sec.~\ref{summary}.

\section{Theoretical framework} \label{theory}

\subsection{Lagrangian density and Hamiltonian}

The starting point of the CDFT is a Lagrangian density where nucleons are described as Dirac spinors with point-coupling or meson-exchange interactions~\cite{Meng2016Book}.
Taking the nonlinear point-coupling version as an example, the Lagrangian density reads
\begin{equation}\label{lagrangian}
\begin{aligned}
\mathcal L = & \bar\psi(i\gamma_\mu \partial^\mu -M)\psi-\frac{1}{2}\alpha_S(\bar\psi\psi)(\bar\psi\psi)-\frac{1}{2}\alpha_V(\bar\psi\gamma_\mu\psi)(\bar\psi\gamma^\mu\psi)-\frac{1}{2}\alpha_{TV}(\bar\psi\vec\tau\gamma_\mu\psi)(\bar\psi\vec\tau\gamma^\mu\psi)\\
&-\frac{1}{2}\alpha_{TS}(\bar\psi\vec\tau\psi)(\bar\psi\vec\tau\psi)-\frac{1}{3}\beta_S(\bar\psi \psi)^3-\frac{1}{4}\gamma_S(\bar \psi \psi)^4-\frac{1}{4}\gamma_V[(\bar\psi\gamma_\mu\psi)(\bar\psi\gamma^\mu\psi)]^2\\
&-\frac{1}{2}\delta_S\partial_\nu(\bar\psi\psi)\partial^\nu(\bar\psi\psi)
-\frac{1}{2}\delta_V\partial_\nu(\bar\psi\gamma_\mu\psi)\partial^\nu(\bar\psi\gamma^\mu\psi)-\frac{1}{2}\delta_{TV}\partial_\nu(\bar\psi\vec\tau\gamma_\mu\psi)\partial^\nu(\bar\psi\vec\tau\gamma^\mu\psi)\\
&-\frac{1}{2}\delta_{TS}\partial_\nu(\bar\psi\vec\tau\psi)\partial^\nu(\bar\psi\vec\tau\psi)-\frac{1}{4}F^{\mu\nu}F_{\mu\nu}-e\bar\psi\gamma^\mu \frac{1-\tau_3}{2}A_\mu \psi,
\end{aligned}
\end{equation}
where $M$ is the nucleon mass, $e$ is the charge unit, $A_\mu$ and $F_{\mu\nu}$ are respectively the four-vector potential and field strength tensor of the electromagnetic field.
Here $\alpha$, $\beta$, $\gamma$, and $\delta$ represent the coupling constants for different channels with the subscripts $S$, $V$, and $T$ standing for the scalar, vector, and isovector couplings, respectively.
The isovector-scalar channel including $\alpha_{TS}$ and $\delta_{TS}$ terms is usually neglected since the isovector-scalar interaction does not improve the description of nuclear ground-state properties~\cite{Burvenich2002PRC,Zhao2010PRC}.

From the Lagrangian density in Eq.~\eqref{lagrangian}, the Hamiltonian density can be obtained via the Legendre transformation,
\begin{equation}
\mathcal{H}=\frac{\partial\mathcal{L}}{\partial\dot{\phi}_i}\dot{\phi}_i-\mathcal{L},
\end{equation}
in which $\phi_i$ represents the nucleon or photon field.
Then, the total Hamiltonian reads
\begin{equation}
\begin{aligned}
H= & \int d^3r\mathcal{H}\\
   =&\int d^3r\{\psi^\dagger[\boldsymbol{\alpha}\cdot\boldsymbol{p}+M]\psi\\
   &+\frac{1}{2}\alpha_{S}(\bar{\psi}\psi)(\bar{\psi}\psi)+\frac{1}{2}\alpha_{V}(\bar{\psi}\gamma_{\mu}\psi)(\bar{\psi}\gamma^{\mu}\psi)+\frac{1}{2}\alpha_{TV}(\bar{\psi}\vec{\tau}\gamma_{\mu}\psi)(\bar{\psi}\vec{\tau}\gamma^{\mu}\psi)\\
   &+\frac{1}{3}\beta_S(\bar{\psi}\psi)^3+\frac{1}{4}\gamma_S(\bar{\psi}\psi)^4+\frac{1}{4}\gamma_V[(\bar{\psi}\gamma_\mu\psi)(\bar{\psi}\gamma^\mu\psi)]^2\\
   &+\frac{1}{2}\delta_{S}(\bar{\psi}\psi)\triangle(\bar{\psi}\psi)+\frac{1}{2}\delta_{V}(\bar{\psi}\gamma_{\mu}\psi)\triangle(\bar{\psi}\gamma^{\mu}\psi)+\frac{1}{2}\delta_{TV}(\bar{\psi}\vec{\tau}\gamma_{\mu}\psi)\triangle(\bar{\psi}\vec{\tau}\gamma^{\mu}\psi)\\
   &+\bar{\psi}e\gamma^\mu A_\mu\psi+\frac{1}{2}A_\mu\triangle A^\mu\}.
\label{Hami}
\end{aligned}
\end{equation}

\subsection{The RHB model}

The physics of weakly bound nuclei necessitates a unified and self-consistent treatment of the mean field and pairing correlations.
This has led to the formulation and development of the RHB model~\cite{Kucharek1991ZPA,Ring1996PPNP}, in which the nuclear ground-state wave function is assumed as a quasiparticle vacuum
\begin{equation}
|\Phi\rangle= \prod_k \beta_k |-\rangle,
 \label{eegs}
\end{equation}
with $|-\rangle$ being the bare vacuum and $\beta_k$ the quasiparticle annihilation operator.
The quasiparticle operators $\beta_k^\dagger$ and $\beta_k$ are defined by a unitary Bogoliubov transformation from particle operators $c_l^\dagger$ and $c_l$ of an arbitrary complete and orthogonal basis,
\begin{equation}
\beta_k^\dagger = \sum_l (U_{lk} c_l^\dagger + V_{lk}c_l),
\end{equation}
where $U$ and $V$ are quasiparticle wave functions.
It can be proved that the quasiparticle vacuum $|\Phi\rangle$ can be uniquely determined by the density matrix $\rho$ and the pairing tensor $\kappa$,
\begin{align}
\rho & =  V^*V^T, \\
\kappa & = V^*U^T.
\end{align}

With the Bogoliubov transformation, the Dirac spinor field $\psi$ and the Hamiltonian in Eq.~\eqref{Hami} can be quantized.
An energy density functional can then be derived from the expectation of the Hamiltonian with respect to the Bogoliubov vacuum in Eq.~\eqref{eegs}, and the variation of the energy density functional with respect to the generalized density matrix~\cite{Peter1980Book},
\begin{equation}
\mathscr{R}=\begin{pmatrix}\rho&\kappa\\-\kappa^*&1-\rho^*\end{pmatrix},
\end{equation}
yields the RHB equations for quasiparticles,
\begin{equation}\label{RHB}
\left(\begin{matrix}
\hat h_D-\lambda_\tau & \hat \Delta \\
-\hat \Delta^* &-\hat h_D^*+\lambda_\tau
\end{matrix}\right)\left(\begin{matrix}
U_k\\
V_k
\end{matrix}\right)=E_k\left(\begin{matrix}
U_k\\
V_k
\end{matrix}\right),
\end{equation}
where $\hat  h_D$ is the Dirac Hamiltonian, $\lambda_\tau$ is the Fermi energy for
neutrons or protons ($\tau=n$ or $p$), $\hat\Delta$ is the pairing potential, and $E_k$ is the quasiparticle energy.

The Dirac Hamiltonian in the coordinate space is
\begin{equation}\label{hD}
h_D(\bm{r})=\bm{\alpha}\cdot\bm{p}+V(\bm{r})+\beta[M+S(\bm{r})],
\end{equation}
with the scalar and vector potentials
\begin{align}
S(\bm r) & =  \alpha_S \rho_S + \beta_S \rho^2_S + \gamma_S \rho^3_S + \delta_S \Delta\rho_S,\label{S} \\
V(\bm r) & = \alpha_V \rho_V + \gamma_V \rho^3_V + \delta_V \Delta\rho_V + e A^0 + \alpha_{TV}\tau_3\rho_3 +\delta_{TV}\tau_3\Delta\rho_3,\label{V}
\end{align}
constructed by various densities
\begin{align}\label{densities}
&\rho_S(\bm r)=\sum_{k>0} V_k^\dag (\bm r)\gamma_0 V_k (\bm r),\\
&\rho_V(\bm r)=\sum_{k>0} V_k^\dag (\bm r) V_k (\bm r),\\
&\rho_3(\bm r)=\sum_{k>0} V_k^\dag (\bm r)\tau_3 V_k (\bm r).
\end{align}
Here $k>0$ means that the summation runs over the positive-energy quasiparticle states in the Fermi sea.

The pairing potential reads
\begin{equation}\label{Delta}
\Delta(\bm{r}_1 s_1 p_1,\bm{r}_2 s_2 p_2)=\sum_{s_1'p_1's_2'p_2'}V^{pp}(\bm{r}_1,\bm{r}_2;s_1p_1,s_2p_2,s_1'p_1',s_2'p_2')\times
\kappa(\bm{r}_1 s_1' p_1',\bm{r}_2 s_2' p_2'),
\end{equation}
where $s$ represents the spin degree of freedom, $p$ denotes the upper or lower component of the Dirac spinor, $V^{pp}$ is the pairing interaction, and $\kappa$ is the pairing tensor~\cite{Peter1980Book}.
In the CDFT, a zero-range density-dependent force and a finite-range Gogny or separable force are usually adopted in the pairing channel~\cite{Meng1998NPA,Meng1998PRC,Long2010PRC,Geng2022PRC,Tian2009PLB,Niksic2014CPC}.
Following the DRHBc theory~\cite{Zhou2010PRC(R),Li2012PRC}, the TRHBc theory employs the zero-range density-dependent pairing force,
\begin{equation}\label{pair}
V^{pp}(\bm r_1,\bm r_2)= V_0 \frac{1}{2}(1-P^\sigma)\delta(\bm r_1-\bm r_2)\left(1-\frac{\rho(\bm r_1)}{\rho_{\mathrm{sat}}}\right),
\end{equation}
where $V_0$ is the pairing strength, $\rho_{\mathrm{sat}}$ represents the saturation density of nuclear matter, and $(1-P^\sigma)/2$ projects onto the spin $S=0$ component.
In this case, both the pairing potential in Eq.~(\ref{Delta}) and the pairing tensor are local~\cite{Li2012PRC}.

\subsection{Triaxially deformed nuclei}

To investigate exotic structures in triaxially deformed nuclei, it is crucial to account for triaxial degrees of freedom.
In the TRHBc framework, various potentials and densities are expanded in spherical harmonics, which naturally accommodate triaxial deformation.
This expansion is well justified, as any three-dimensional function can be expressed as a sum of spherical harmonics with radial functions as expansion coefficients~\cite{Arfken2012Book},
\begin{equation}\label{legendre}
      f(\bm r)  =  \sum_{\lambda\mu} f_{\lambda\mu}(r)Y_{\lambda\mu}(\theta,\varphi),
\end{equation}
where
\begin{equation}
  f_{\lambda\mu}(r)  = \int d\Omega Y_{\lambda\mu}^*(\theta,\varphi)f(\bm r).
\end{equation}
For the spherical harmonic expansion in Eq.~(\ref{legendre}), limitations on $\lambda$ and $\mu$ can be obtained by symmetry~\cite{Xiang2023Symmetry}.
Assuming the spatial reflection symmetry and mirror symmetries with respect to the $xy$, $xz$, and $yz$ planes, $\lambda$ and $\mu$ are restricted to be even numbers, and the $+\mu$ component equals to the $-\mu$ one, i.e., $f_{\lambda\mu}(r)=f_{\lambda-\mu}(r)$.
Details of the symmetry analysis are given in Appendix \ref{symm}.
Finally, the expansion in Eq.~(\ref{legendre}) in the TRHBc theory is reduced to
\begin{equation}\label{SHF}
\begin{split}
f(\bm r)& =\sum_{\lambda=0,2,...}f_{\lambda0}(r)Y_{\lambda0}(\Omega)+\sum_{\lambda=2,4,...}^{\mu=2,...,\lambda} f_{\lambda \mu}(r)[Y_{\lambda\mu}(\Omega)+Y_{\lambda-\mu}(\Omega)]\\
& = \sum_{\lambda=0,2,...}f_{\lambda0}(r)Y_{\lambda0}(\Omega)+2\sum_{\lambda=2,4,...}^{\mu=2,...,\lambda} f_{\lambda \mu}(r)\text{Re}[Y_{\lambda\mu}(\Omega)].
\end{split}
\end{equation}

\subsection{Contribution of the continuum}

For exotic nuclei with Fermi energies near the continuum threshold, pairing correlations can scatter nucleons from bound to resonant states, which may result in diffuse densities.
Such continuum effects are essential for describing both spherical and axially deformed halo nuclei~\cite{Meng1996PRL,Zhou2010PRC(R)}.
The DRHBc theory incorporates the possible large spatial extension of exotic nuclei by employing the DWS basis~\cite{Zhou2003PRC}, the wave function of which has an appropriate asymptotic behavior at large $r$.
In the TRHBc theory, the triaxial RHB equations are likewise solved in the DWS basis,
\begin{equation}\label{WFWS}
|n\kappa m\rangle =i^p \frac{R_{n\kappa}(r,p)}{r}\mathcal{Y}_{j m}^{l(p)}(\Omega,s),
\end{equation}
where $r$ represents the radial coordinate, $s$ denotes the spin, and $p =1$ or $2$ is the index for the upper or lower component.
The DWS basis is characterized by its quantum numbers $n$, $\kappa$, and $m$, in which $n$ is the node number of the radial wave function $R_{n\kappa}$, $\kappa$ is given by the parity $\pi$ and the total angular momentum $j$, $\kappa=(-1)^{j+l+1/2}(j+1/2)$, and $m$ is the third component of $j$.
Further details about the DWS basis can be found in Appendix \ref{DWS}.

With a set of DWS bases, solving the RHB equation is equivalent to the diagonalization of the RHB matrix.
Because of the conserved spatial reflection symmetry, the RHB matrix can be decomposed into the parity $\pi = +$ and $\pi = -$ blocks.
In each block, the diagonalization,
\begin{equation}\label{RHBm}
\left( \begin{matrix}
\mathcal{A}-\lambda_\tau & \mathcal{B} \\
\mathcal{B}^\dag & -\mathcal{A}^*+\lambda_\tau
\end{matrix}\right)\left( \begin{matrix}
\mathcal{U}_k \\
\mathcal{V}_k
\end{matrix}\right)=
E_k\left( \begin{matrix}
\mathcal{U}_k \\
\mathcal{V}_k
\end{matrix}\right),
\end{equation}
leads to the quasiparticle energy $E_k$ and the eigenvectors
\begin{equation}
\mathcal{U}_k=(u_{k,(n\kappa m)}),~~\mathcal{V}_k=(v_{k,(n\kappa m)}),
\end{equation}
which correspond to the expansion coefficients of quasiparticle wave functions in the DWS basis.
The matrix elements in Eq. (\ref{RHBm}) are
\begin{align}
& \mathcal{A}= h_{D(n\kappa m)(n'\kappa' m')} = \langle n\kappa m|h_D|n'\kappa'm'\rangle,\\
& \mathcal{B}= \Delta_{(n\kappa m)(n'\kappa' m')} = \langle n\kappa m|\Delta|\overline{n'\kappa'm'}\rangle,
\end{align}
where $|\overline{n'\kappa'm'}\rangle$ is the time-reversal state of $|n'\kappa'm'\rangle$.
Details for the calculation of RHB matrix elements are given in Appendix \ref{ME}.

\subsection{Blocking effects for odd system}

For odd-mass and odd-odd nuclei, the blocking effects of the unpaired nucleon(s) should be taken into account.
Starting from the ground state $|\Phi\rangle$ in Eq.~\eqref{eegs} of an even-even nucleus, an odd-mass nucleus containing an unpaired particle can be described by a one-quasiparticle state,
\begin{equation}
|\Phi_1\rangle=\beta_{k_b}^\dagger |\Phi\rangle=\beta_{k_b}^\dagger\prod_k\beta_k |0\rangle,
\end{equation}
where $\beta_{k_b}^\dagger$ denotes the creation operator for the blocked quasiparticle state $k_b$.
The one-quasiparticle state $|\Phi_1\rangle$ is the vacuum with respect to the set of quasiparticle operators $(\beta_{1},\ldots,\beta_{k_{b}}^{\dagger},\ldots)$.
Thus, the blocking can be implemented by the exchange of $\beta_{k_b}\leftrightarrow\beta_{k_b}^{\dagger}$.
This exchange corresponds to the exchange of the column vectors $(V_{k_b}^*,U_{k_b}^*)\leftrightarrow(U_{k_b},V_{k_b})$ and the energy shift $E_{k_b}\leftrightarrow -E_{k_b}$.
The blocking effects in an odd-odd nucleus can be treated in a similar way.
The above strict treatment would break the time reversal symmetry, and as a result nuclear currents and time-odd fields emerge~\cite{Pan2024PLB}.

Following Refs.~\cite{Li2012CPL,Pan2022PRC}, the equal filling approximation (EFA)~\cite{Perez-Martin2008PRC} is adopted to incorporate the blocking effects in the present TRHBc theory.
In the EFA, two configurations with the quasiparticle state $k_b$ and its conjugate state respectively blocked are averaged in a statistical way, which conserves the time reversal symmetry.
It therefore avoids dealing with the complicated currents and time-odd fields, and the density matrix and the pairing tensor read~\cite{Li2012CPL}
\begin{align}
\rho' & = \rho +  \frac{1}{2}(U_{k_b}U_{k_b}^{*T} - V_{k_b}^*V_{k_b}^T), \\
\kappa' & = \kappa - \frac{1}{2}(U_{k_b}V_{k_b}^{*T} + V_{k_b}^*U_{k_b}^T).
\end{align}

\subsection{Physical observables}

After solving self-consistently the RHB equation, physical quantities including the total energy, root-mean-square (rms) radii, and deformation parameters can be calculated.
The total energy of a nucleus is
\begin{equation}
	E_{\mathrm{RHB}} =E_{\mathrm{nucleon}} + E_\mathrm{pair} + E_S + E_V + E_{\mathrm{c.m.}} + E_{\mathrm{rot}}.
\label{Etot}
\end{equation}
The nucleon energy $E_{\mathrm{nucleon}}$ reads
\begin{equation}
	E_{\mathrm{nucleon}} = \sum_{k>0} (\lambda_\tau - E_k) v_k^2 - 2E_{\mathrm{pair}} ,
\end{equation}
with
\begin{equation}
	v_k^2=\int d^3 \bm{r} V_k^\dag(\bm{r}) V_k(\bm{r}) .
\end{equation}
For an odd-mass nucleus with the quasiparticle state $k_b$ blocked, the nucleon energy becomes
\begin{equation}
	E_{\mathrm{nucleon}} = \sum_{k>0} (\lambda_\tau - E_k) v_k^2 + (\lambda_\tau + E_{k_b})u_{k_b}^2 - (\lambda_\tau - E_{k_b})v_{k_b}^2 - 2E_{\mathrm{pair}},
\end{equation}
in which $u_k^2 = 1-v_k^2$.
The pairing energy $E_{\mathrm{pair}}$, with the zero-range pairing force, is calculated by
\begin{equation}
	E_{\mathrm{pair}} = -\frac{1}{2} \int d^3 \bm{r} \kappa(\bm{r}) \Delta(\bm{r}).
\end{equation}
The scalar and vector parts of the potential energy are respectively
\begin{align}
E_S &= -  \int \mathrm{d}^3 \bm r \left(\frac{1}{2}\alpha_S \rho_S^2 + \frac{2}{3}\beta_S \rho^3_S + \frac{3}{4}\gamma_S \rho^4_S + \frac{1}{2}\delta_S \rho_S \triangle\rho_S \right), \\
E_V &=  - \int \mathrm{d}^3 \bm r \left(\frac{1}{2}\alpha_V \rho_V^2 + \frac{1}{2}\alpha_{TV}\rho_3^2 + \frac{3}{4}\gamma_V \rho^4_V + \frac{1}{2} \delta_V \rho_V \triangle\rho_V +\frac{1}{2}\delta_{TV}\rho_3 \triangle\rho_3 +\frac{1}{2} \rho_p e A^0\right).
\end{align}
The center-of-mass and rotational correction energies can be calculated respectively by
\begin{equation}
\label{Ecm}
	E_{\mathrm{c.m.}} = -\frac{ \braket{\hat {\bm{P}}^2} }{2MA},
\end{equation}
and
\begin{equation}
\label{Erot}
	E_{\mathrm{rot}} = -\frac{ \braket{\hat{\bm{J}}^2} }{2 \mathscr{I}},
\end{equation}
where $\hat{\bm{P}} = \sum_i^A \hat{\bm{p}}_i$ is the total momentum in the center-of-mass frame~\cite{Bender2000EPJA}, $A$ the mass number, $\hat{\bm{J}}$ the total angular momentum, and $\mathscr{I}$ the moment of inertia that can be evaluated using the Inglis-Belyaev formula~\cite{Peter1980Book}.

The rms radius is calculated by
\begin{equation}\label{radius}
R=  \sqrt{\int r^2\rho_v(\bm r) d^3 \bm r},
\end{equation}
where $\rho_v$ is the normalized vector density of neutrons, protons, or nucleons.
The charge rms radius $R_{\mathrm{ch}}$ is calculated based on the proton rms radius $R_p$,
\begin{equation}
	R_{\mathrm{ch}}= \sqrt{R_{p}^2 + 0.64~\mathrm{fm}^2}.
\end{equation}

The intrinsic quadrupole moments are calculated as
\begin{align}
& Q_{20}  = \int r^2Y_{20}\rho_V(\bm r) d^3 \bm r,\\
& Q_{22}  = \frac{1}{2}\int r^2(Y_{22} + Y_{2-2})\rho_V(\bm r) d^3 \bm r.
\end{align}
Then, the deformation parameters $\beta$ and $\gamma$ can be calculated by
\begin{equation}
\beta =\frac{4\pi}{3Ar_0^2} \sqrt{Q_{20}^2+2Q_{22}^2},~~\gamma=\arctan(\sqrt{2}\frac{Q_{22}}{Q_{20}} ).
\end{equation}
where $r_0 = 1.2 A^{1/3}$~fm is the empirical matter radius of a nucleus.

\section{Numerical details}\label{numerical}

For a given relativistic density functional and a pairing interaction, the TRHBc calculations can be performed for any nuclei without introducing any free parameters.
As the RHB equations are solved in the DWS basis, the box size and the mesh size as well as the energy and angular momentum cutoff for the basis should be checked for convergence.
For completeness, basis states from both the Fermi and Dirac seas are included, typically with equal numbers~\cite{Zhou2003PRC,Zhou2010PRC(R)}.
The spherical harmonic expansion in Eq.~\eqref{SHF} truncated at finite order in practice should also be checked for convergence.

In Ref.~\cite{Zhou2003PRC}, it was shown that the solutions of relativistic Hartree equations in the DWS basis with the box size $R_{\mathrm{box}}=20$ fm and the mesh size $\Delta r=0.1$ fm accurately reproduce the results obtained from the shooting method.
It has been further confirmed in Ref.~\cite{Zhang2020PRC} that $R_{\mathrm{box}}=20$ fm and $\Delta r=0.1$ fm are enough to provide converged results in the DRHBc calculations for both light and heavy nuclei.
In the TRHBc calculations for the triaxial nucleus $^{86}$Ge (whose deformation is illustrated below), we found that $R_{\mathrm{box}}=20$ fm and $\Delta r=0.1$ fm lead to a satisfactory accuracy, with an uncertainty less than $0.01\%$ in its binding energy.
Therefore, a box size $R_{\mathrm{box}}$ of 20 fm and a mesh size $\Delta r$ of 0.1 fm, the same as those used in the RCHB and DRHBc mass table calculations~\cite{Xia2018ADNDT,Zhang2022ADNDT}, are adopted in the present TRHBc calculations.

\subsection{Energy cutoff}

\begin{figure}[htbp]
  \centering
  \includegraphics[scale=0.4,angle=0]{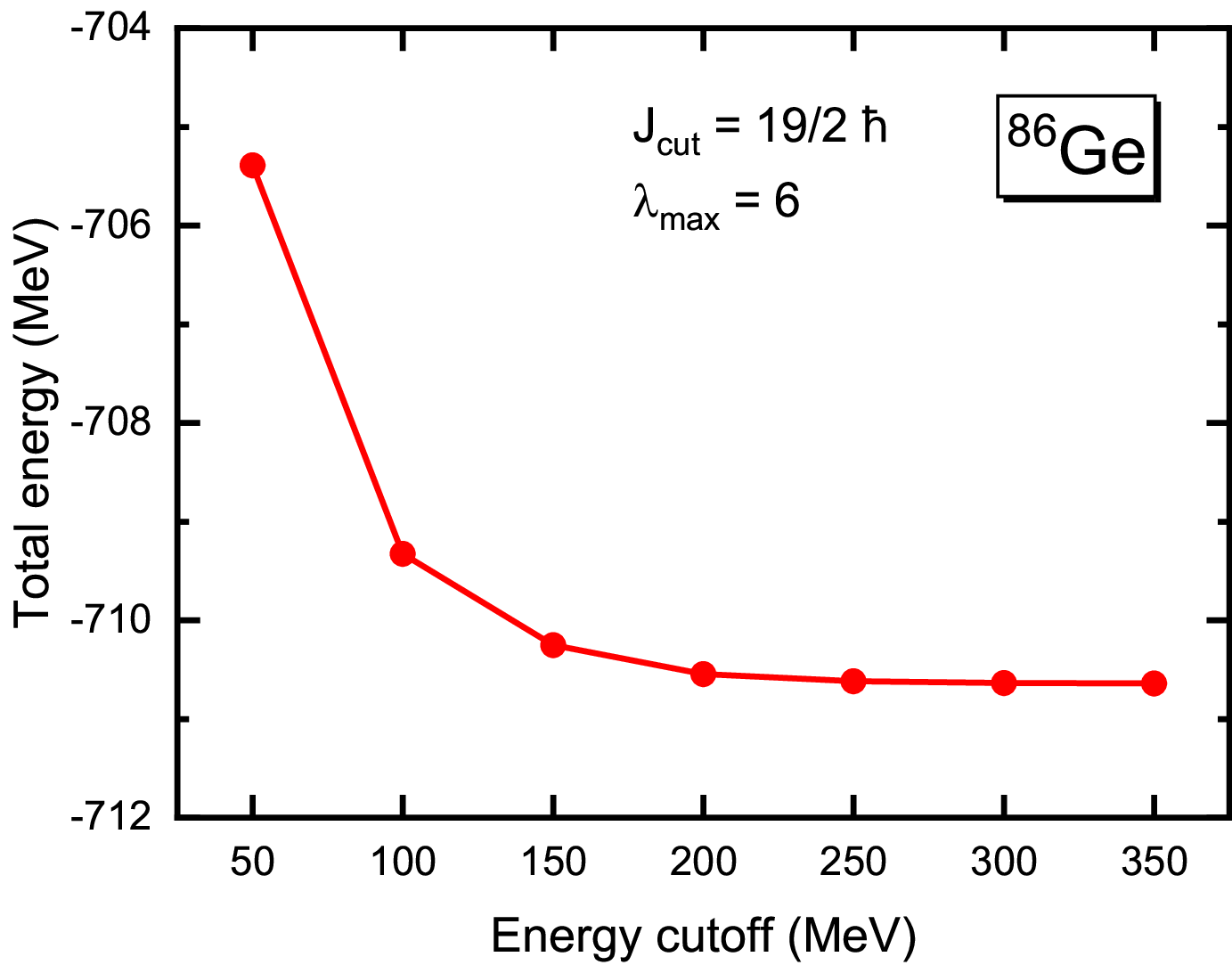}
  \caption{Total energy as a function of the energy cutoff $E_{\mathrm{cut}}$ for the triaxial nucleus $^{86}$Ge calculated by the TRHBc theory with density functional PC-PK1. Here an angular momentum cutoff $J_{\mathrm{cut}}=19/2$ $\hbar$ and a spherical harmonic expansion truncation $\lambda_{\mathrm{max}}=6$ are used, and the pairing correlation is neglected.}
\label{fig1}
\end{figure}

In Ref.~\cite{Zhang2020PRC}, numerical checks for the energy cutoff have been performed, and $E_{\mathrm{cut}} = 300$ MeV has been suggested for the DRHBc mass table calculation.
Here, we examine the convergence of the total energy with respect to $E_{\mathrm{cut}}$ in Fig. \ref{fig1} for $^{86}$Ge.
In Fig. \ref{fig1}, the total energy gradually converges as $E_{\mathrm{cut}}$ increases.
Quantitatively, the difference between the results obtained with $E_{\mathrm{cut}} = 300$ MeV and $E_{\mathrm{cut}} = 350$ MeV is less than 0.01 MeV.
Therefore, an energy cutoff of $E_{\mathrm{cut}} = 300$ MeV suffices to yield converged results in the study using the TRHBc theory.

\subsection{Angular momentum cutoff}

\begin{figure}[htbp]
  \centering
  \includegraphics[scale=0.4,angle=0]{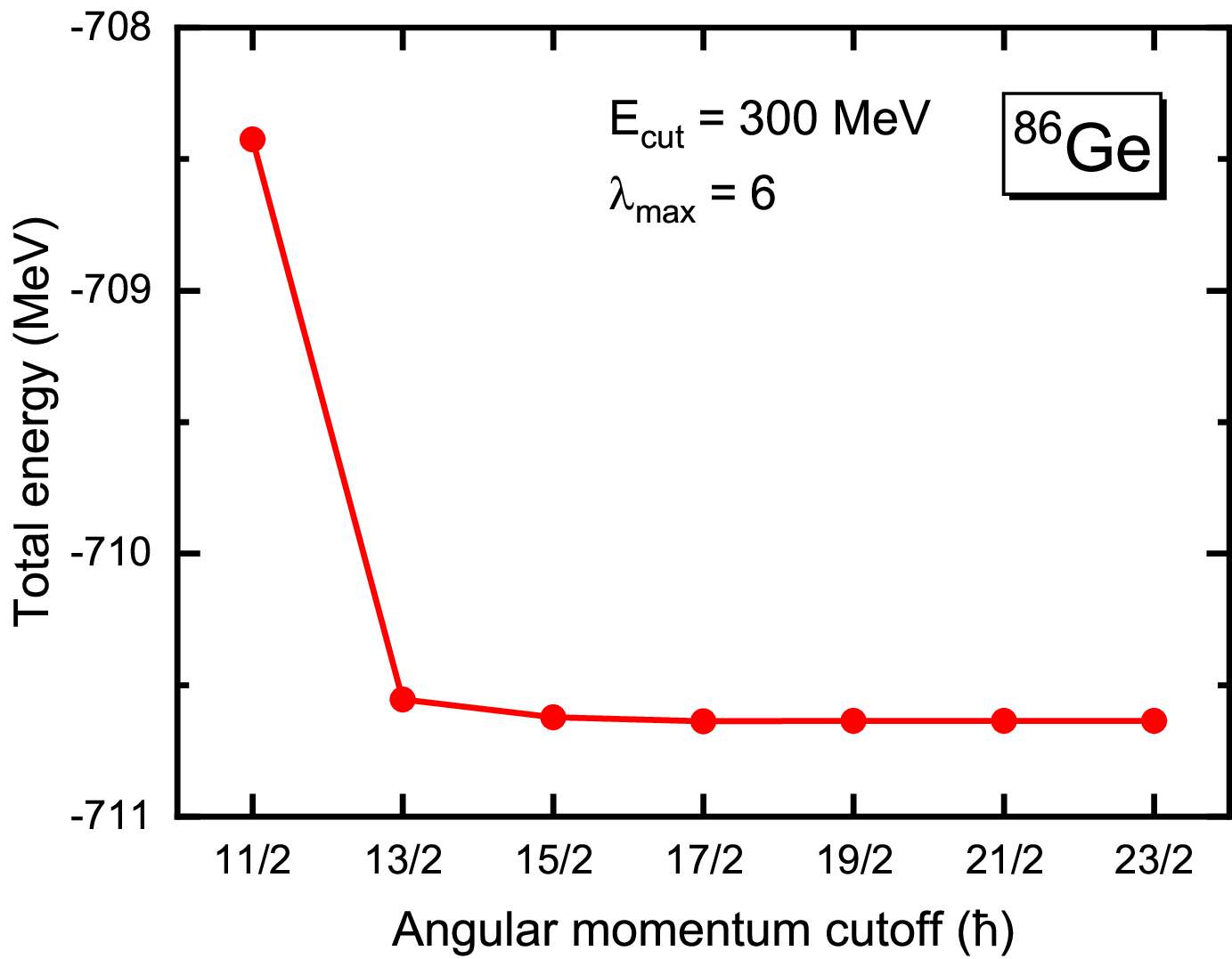}
  \caption{Total energy as a function of the angular momentum cutoff $J_{\mathrm{cut}}$ for the triaxial nucleus $^{86}$Ge calculated by the TRHBc theory with density functional PC-PK1. Here an energy cutoff $E_{\mathrm{cut}}=300$ MeV and a spherical harmonic expansion truncation $\lambda_{\mathrm{max}}=6$ are used, and the pairing correlation is neglected.}
\label{fig2}
\end{figure}

In the RCHB mass table calculation~\cite{Xia2018ADNDT}, an angular momentum cutoff of $J_{\mathrm{cut}}=19/2~\hbar$ was employed, and this cutoff has been demonstrated to yield converged results for light and medium-heavy nuclei in DRHBc calculations~\cite{Zhang2020PRC}.
To assess the convergence with respect to $J_{\mathrm{cut}}$ in TRHBc calculations, we conduct a check for $^{86}$Ge, as illustrated in Fig. \ref{fig2}.
The total energy of $^{86}$Ge exhibits negligible variation for $J_{\mathrm{cut}}\ge 15/2$ $\hbar$.
The quantitative difference between results obtained with $J_{\mathrm{cut}} = 19/2$, $21/2$, and $23/2$ $\hbar$ is less than $0.001$ MeV.
Hence, an angular momentum cutoff of $J_{\mathrm{cut}}=19/2$ $\hbar$ is sufficient to achieve converged results in TRHBc calculations.

\subsection{Spherical harmonic expansion truncation}

\begin{figure}[htbp]
  \centering
  \includegraphics[scale=0.4,angle=0]{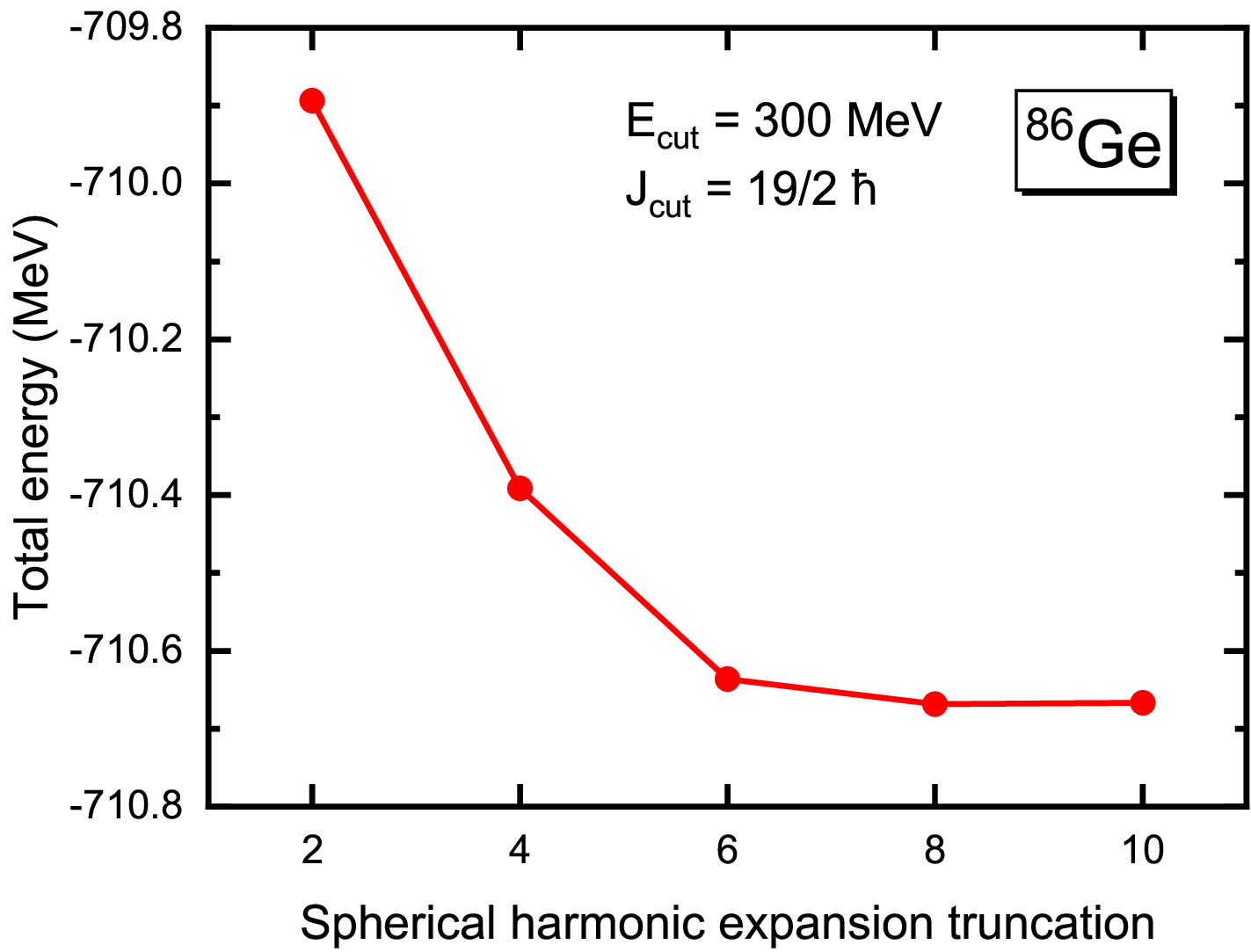}
  \caption{Total energy as a function of the spherical harmonic expansion $\lambda_{\mathrm{max}}$ for the triaxial nucleus $^{86}$Ge calculated by the TRHBc theory with density functional PC-PK1. Here an energy cutoff $E_{\mathrm{cut}}=300$ MeV and an angular momentum cutoff $J_{\mathrm{cut}}=19/2$ $\hbar$ are used, and the pairing correlation is neglected.}
\label{fig3}
\end{figure}

In the DRHBc theory, the axially deformed densities and potentials are expanded in terms of Legendre polynomials $P_\lambda$, which satisfies the following relationship with the spherical harmonic function,
\begin{equation}
P_\lambda(\cos\theta) = \sqrt{\frac{4\pi}{2\lambda+1}}Y_{\lambda 0}(\theta).
\end{equation}
It has been demonstrated that, for light nuclei and medium-heavy nuclei, truncating the Legendre expansion at $\lambda_{\mathrm{max}}=6$ ensures converged results in DRHBc calculations~\cite{Pan2019IJMPE,Zhang2020PRC}.
Here we perform the convergence check for the spherical harmonic expansion using the triaxial nucleus $^{86}$Ge as an example.
Figure \ref{fig3} displays its total energy as a function of the spherical harmonic expansion truncation $\lambda_{\mathrm{max}}$.
A clear converging trend is observed with increasing $\lambda_{\mathrm{max}}$.
Quantitatively, the difference between total energies calculated with $\lambda_{\mathrm{max}}=6$, $8$, and $10$ is approximately 0.03 MeV, around $0.004\%$ of the total energy.
Hence, the conclusion that truncating at $\lambda_{\mathrm{max}}=6$ leads to converged results remains valid after including triaxiality.

\subsection{Pairing strength}

\begin{figure}[htbp]
  \centering
  \includegraphics[scale=0.4,angle=0]{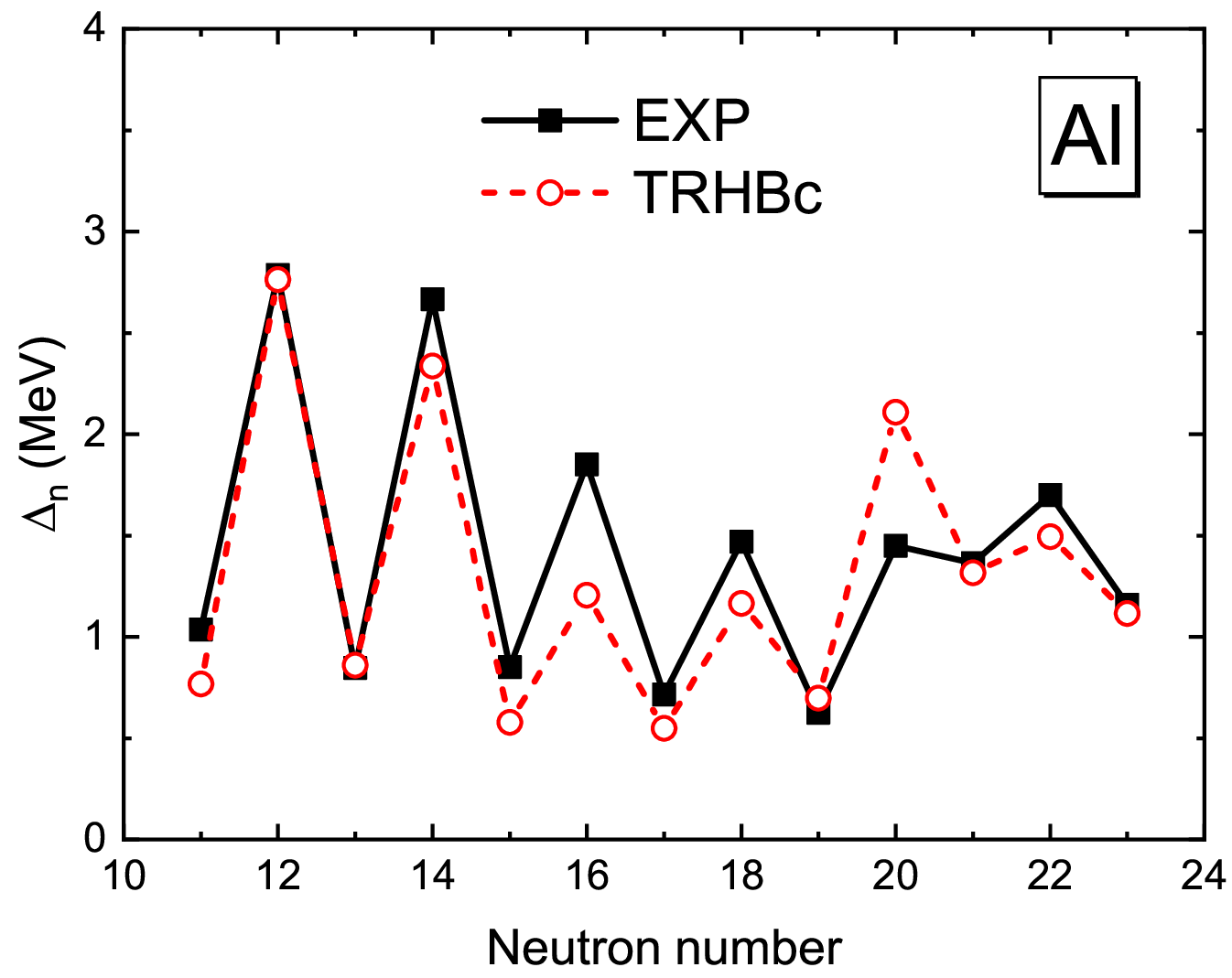}
  \caption{Odd-even mass difference as a function of the neutron number for aluminum isotopes calculated by the TRHBc theory with density functional PC-PK1. Here an energy cutoff $E_{\mathrm{cut}}=300$ MeV, an angular momentum cutoff $J_{\mathrm{cut}}=19/2$ $\hbar$, a spherical harmonic expansion truncation $\lambda_{\mathrm{max}}=6$, and a pairing strength $V_0=-342.5$ MeV fm$^3$ are used. The available data extracted from experimental binding energies are also shown for comparison.}
\label{fig}
\end{figure}

Utilizing $J_{\mathrm{cut}}=19/2$ $\hbar$ and a pairing window of 100 MeV, the RCHB calculations with the PC-PK1 density functional, pairing strength $V_0=-342.5$ MeV fm$^3$, and saturation density $\rho_{\mathrm{sat}}=0.152$ fm$^{-3}$ well reproduce the experimental odd-even mass differences for calcium, tin, lead, and uranium isotopes, as well as $N =20$ and $50$ isotones~\cite{Xia2018ADNDT}.
The odd-even mass differences for calcium and lead isotopes have also been reproduced by DRHBc calculations employing the same numerical details~\cite{Zhang2020PRC}.
To achieve a unified description across spherical, axially deformed, and triaxially deformed nuclei within the TRHBc framework, no further adjustment of the pairing strength is needed.
In Fig.~\ref{fig}, the calculated odd-even mass differences are compared with the available data for aluminum isotopes. It is shown that the TRHBc calculations reproduce the data well, validating the correctness of the pairing strength.

Finally,  the suggested numerical details for TRHBc calculations of light and medium-heavy nuclei are summarized as follows. The box size $R_{\mathrm{box}}= 20$ fm, the mesh size $\Delta r = 0.1$ fm, the energy cutoff $E_{\mathrm{cut}} = 300$ MeV, the angular momentum cutoff $J_{\mathrm{cut}} = 19/2$ $\hbar$, the spherical harmonic expansion truncation $\lambda_{\mathrm{max}} = 6$, the pairing strength $V_0 = -342.5$ MeV fm$^3$, the saturation density $\rho_{\mathrm{sat}}=0.152$ fm$^{-3}$, and the pairing window is 100 MeV.
These ensure the convergence of the numerical results, with no free parameters introduced in present TRHBc calculations.

\subsection{Benchmark calculations}

\begin{figure}[htbp]
  \centering
  \includegraphics[scale=0.4,angle=0]{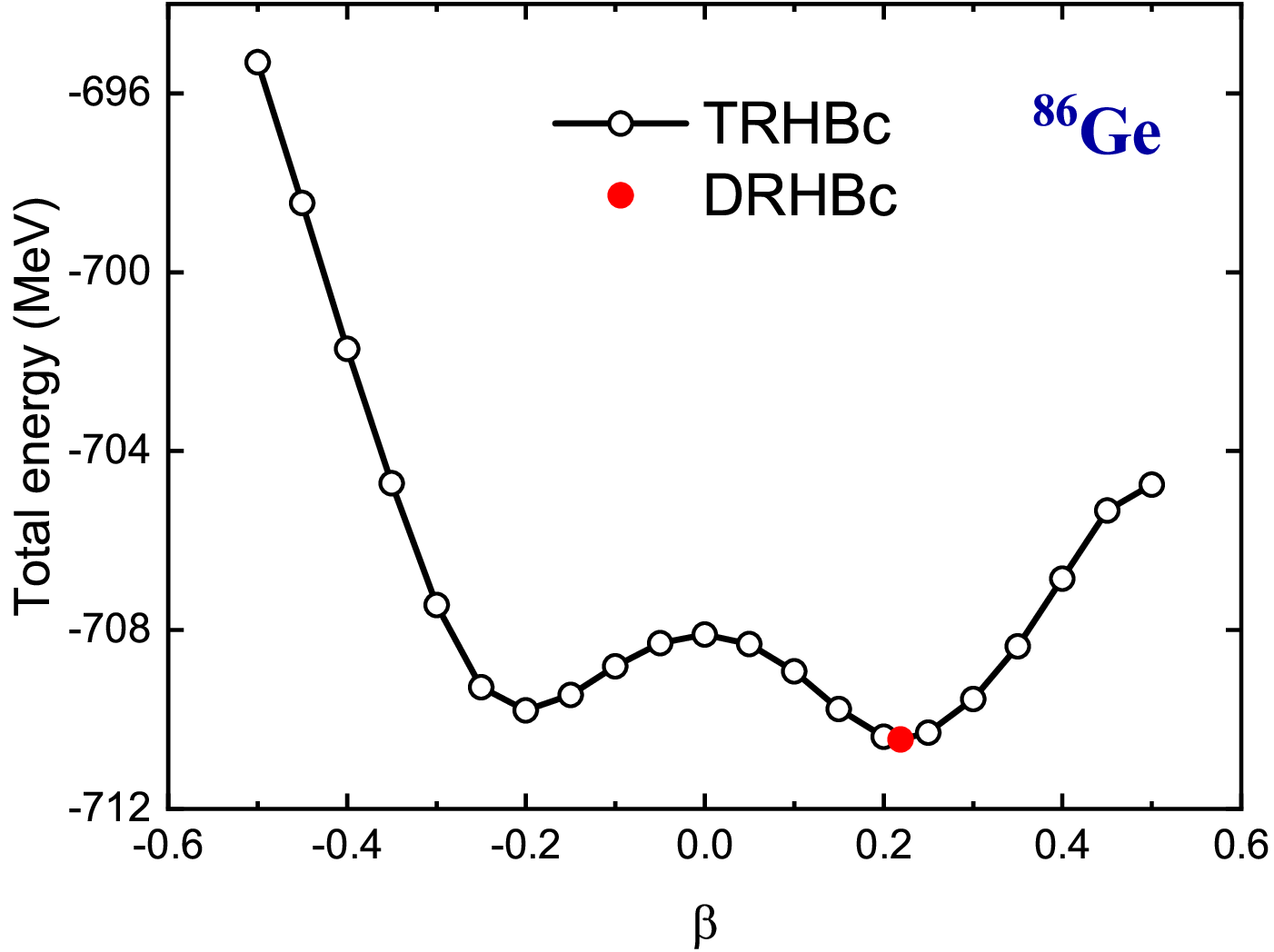}
  \caption{Potential energy curve for $^{86}$Ge from constrained TRHBc calculations with the density functional PC-PK1. The energy and deformation of the prolate local minimum from unconstrained DRHBc calculations are labeled by the filled circle.}
\label{fig4}
\end{figure}

To further benchmark the numerical implementation of the TRHBc theory, the constrained TRHBc calculations are first performed with $\gamma = 0^\circ$ and $60^\circ$ to construct the potential energy curve (PEC) for $^{86}$Ge.
The obtained results are shown in Fig. \ref{fig4}, in comparison with the ground state from the unconstrained DRHBc calculations using the same numerical details.
It can be found that the ground-state energy and deformation from the DRHBc calculations match the minimum point on the prolate side of the PEC, validating the consistence between the TRHBc and DRHBc theories in describing axially deformed systems.

\begin{figure}[htbp]
  \centering
  \includegraphics[scale=0.5,angle=0]{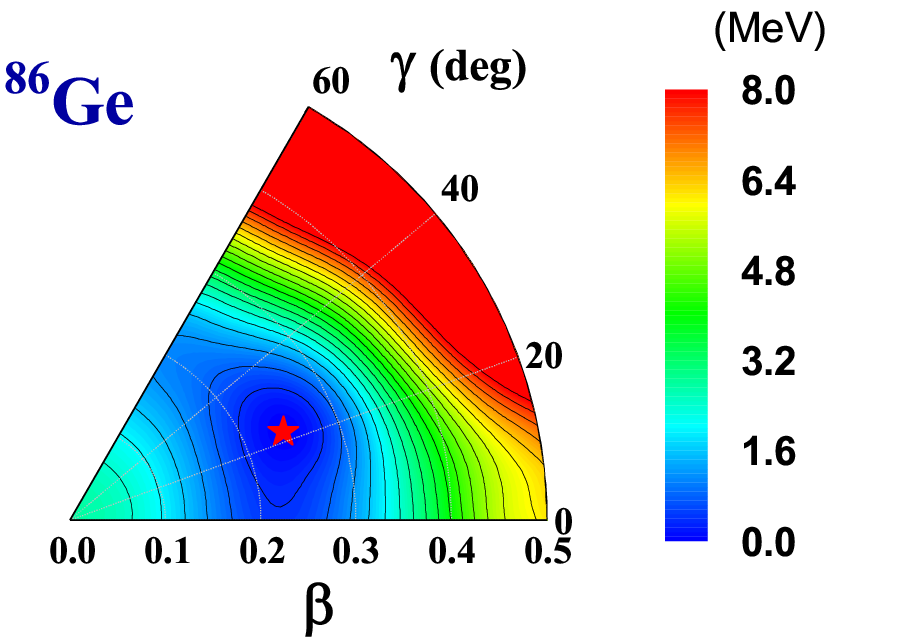}
  \caption{Potential energy surface for $^{86}$Ge in the $\beta$-$\gamma$ plane from constrained TRHBc calculations with the density functional PC-PK1. All energies are normalized with respect to the energy of the absolute minimum (in MeV) indicated by the star. The energy separation between each contour line is 0.4 MeV.}
\label{fig5}
\end{figure}

In Fig. \ref{fig5}, the potential energy surface (PES) for $^{86}$Ge from the constrained TRHBc calculations is illustrated, with the ground-state deformation from unconstrained calculations highlighted by a star.
The ground-state deformation aligns with the absolute minimum of the PES, verifying the self-consistency of the present TRHBc calculations.
The deformation parameters $(\beta,\gamma)$ for the ground state are $(0.24,22^\circ)$, consistent with the values of $(0.24,23^\circ)$ obtained from the TRHB calculations~\cite{Yang2021PRC} employing the separable pairing force~\cite{Tian2009PLB} and the harmonic oscillator basis.
It is also crucial to note that both the prolate and oblate minima on the PEC shown in Fig. \ref{fig4} are actually saddle points on the PES.
Therefore, the inclusion of triaxial deformation degrees of freedom is of paramount significance for achieving a fully microscopic description.

\section{Results and Discussions}\label{results}

In this section, we present and discuss the results for aluminum isotopes from the TRHBc theory.
Ground-state properties including binding energies, rms radii, and deformation parameters for the bound aluminum isotopes calculated with density functionals PC-PK1~\cite{Zhao2010PRC}, NL3$^*$~\cite{Lalazissis2009PLB}, NLSH~\cite{Sharma1993PLB}, and PK1~\cite{Long2004PRC} are listed respectively in Tables \ref{tab1}--\ref{tab4} in Appendix \ref{tables}.
The available data for binding energies~\cite{AME2020(1),AME2020(2),AME2020(3)} and charge radii~\cite{Heylen2021PRC} are also included for comparison.

\subsection{Nuclear binding and size}

In Ref.~\cite{Zhang2023PRC(L2)}, the TRHBc calculated one-neutron separation energies and charge radii have been compared with available data.
The TRHBc calculations yield the correct proton drip-line location at $^{22}$Al~\cite{Wu2021PRC,Campbell2024PRL}, beyond which the lighter isotopes are no longer bound against proton emission.
The one-neutron separation energies~\cite{AME2020(3)} and the odd-even staggering are also well reproduced by the TRHBc calculations, both in trend and in magnitude.
The one-neutron drip-line nucleus is predicted to be $^{43}$Al, the heaviest aluminum isotope observed so far~\cite{Baumann2007Nature}.
Furthermore, the calculated charge radii agree with the data within experimental uncertainties~\cite{Heylen2021PRC}.

As shown in Tables \ref{tab1}--\ref{tab4}, the experimental binding energies are reproduced reasonably well by the TRHBc theory with the four density functionals.
The rms deviations from the data are respectively 1.86, 2.81, 2.23, and 2.08 MeV for PC-PK1, NL3$^*$, NLSH, and PK1.
As one of the most successful density functionals, PC-PK1 provides the best results here.
Its predictive power has been proven by reproducing the masses of heavy nuclei measured at GSI~\cite{Zhao2012Phys.Rev.C64324} and those of superheavy nuclei complied in AME2020~\cite{Zhang2021PRC(L),He2024PRC}.
It is worth mentioning that only the center-of-mass correction energies in Eq.~\eqref{Ecm} have been included in the present TRHBc calculation.
In Refs.~\cite{Zhang2014FOP,Zhang2022ADNDT}, it is shown that the rotational correction energies in Eq.~\eqref{Erot} would improve the PC-PK1 description by more than 1 MeV.
An alternative to Eq. \eqref{Erot} is the collective Hamiltonian method, which evaluates simultaneously the rotational and vibrational correction energies for both well deformed and near spherical nuclei, thereby further refining the description of nuclear masses and nucleon separation energies~\cite{Lu2015PRC,Yang2021PRC,Sun2022CPC}.

\begin{figure}[htbp]
  \centering
  \includegraphics[width=0.5\textwidth]{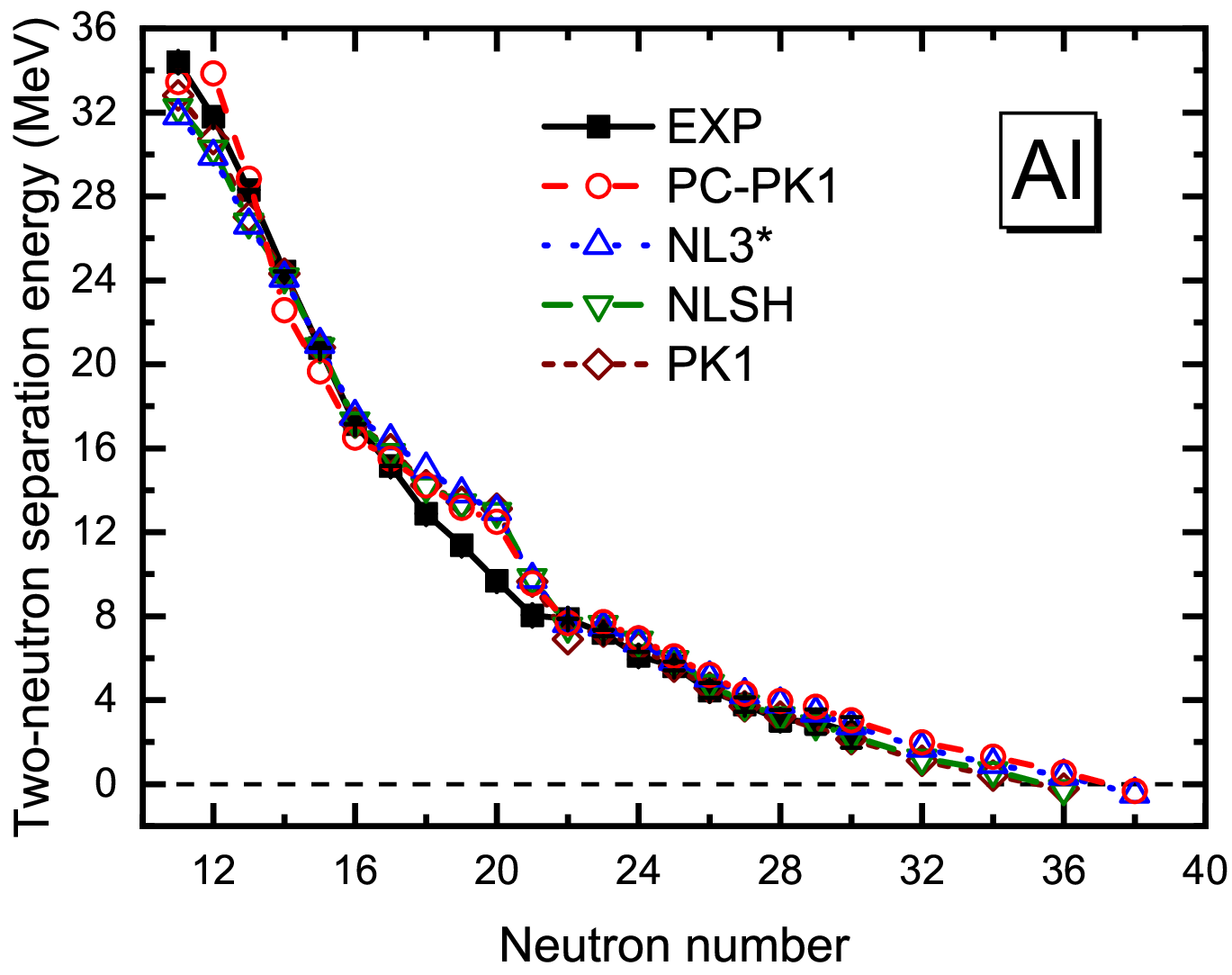}
  \caption{Two-neutron separation energy as a function of the neutron number for aluminum isotopes from TRHBc calculations, in comparison with the AME2020 data~\cite{AME2020(3)}.}
\label{fig6}
\end{figure}

Figure \ref{fig6} shows the two-neutron separation energies $S_{2n}$ calculated by the TRHBc theory, in comparison with the AME2020 data~\cite{AME2020(3)}, including experimental values for $12\le N\le 25$ and empirical ones for others.
The AME empirical data, estimated from the trends in mass surface and all available experimental information~\cite{AME2020(2)}, are generally corroborated by subsequent measurements~\cite{Michimasa2020PRL,Fu2020PRC}.
Overall, the results from different density functionals are in good agreement with the AME2020 data, except for the deviation near $N= 20$.
The calculated $S_{2n}$ displays a sudden drop at $N =20$, which is an indicator of the neutron shell closure~\cite{Zhang2005NPA,Li2014PLB}.
The experimental trend shows a more gradual decrease that does not support a robust $N=20$ shell gap, referred as the island of inversion~\cite{Tripathi2008PRL,Doornenbal2009PRL,Wimmer2010PRL}, the description of which is challenging at purely the mean-field level~\cite{Li2012PRC}.
In contrast, the TRHBc theory correctly captures the collapse of the $N=28$ shell closure in neutron-rich aluminum isotopes.
The two-neutron drip-line nucleus is predicted as $^{49}$Al by PC-PK1 and NL3$^*$, while as $^{47}$Al by NLSH and PK1.
As the experimental neutron drip line has been established only up to the neon isotopic chain~\cite{Ahn2019PRL}, these predictions await future verification.

\subsection{Triaxial deformation effects}

\begin{figure}[htbp]
  \centering
  \includegraphics[width=0.95\textwidth]{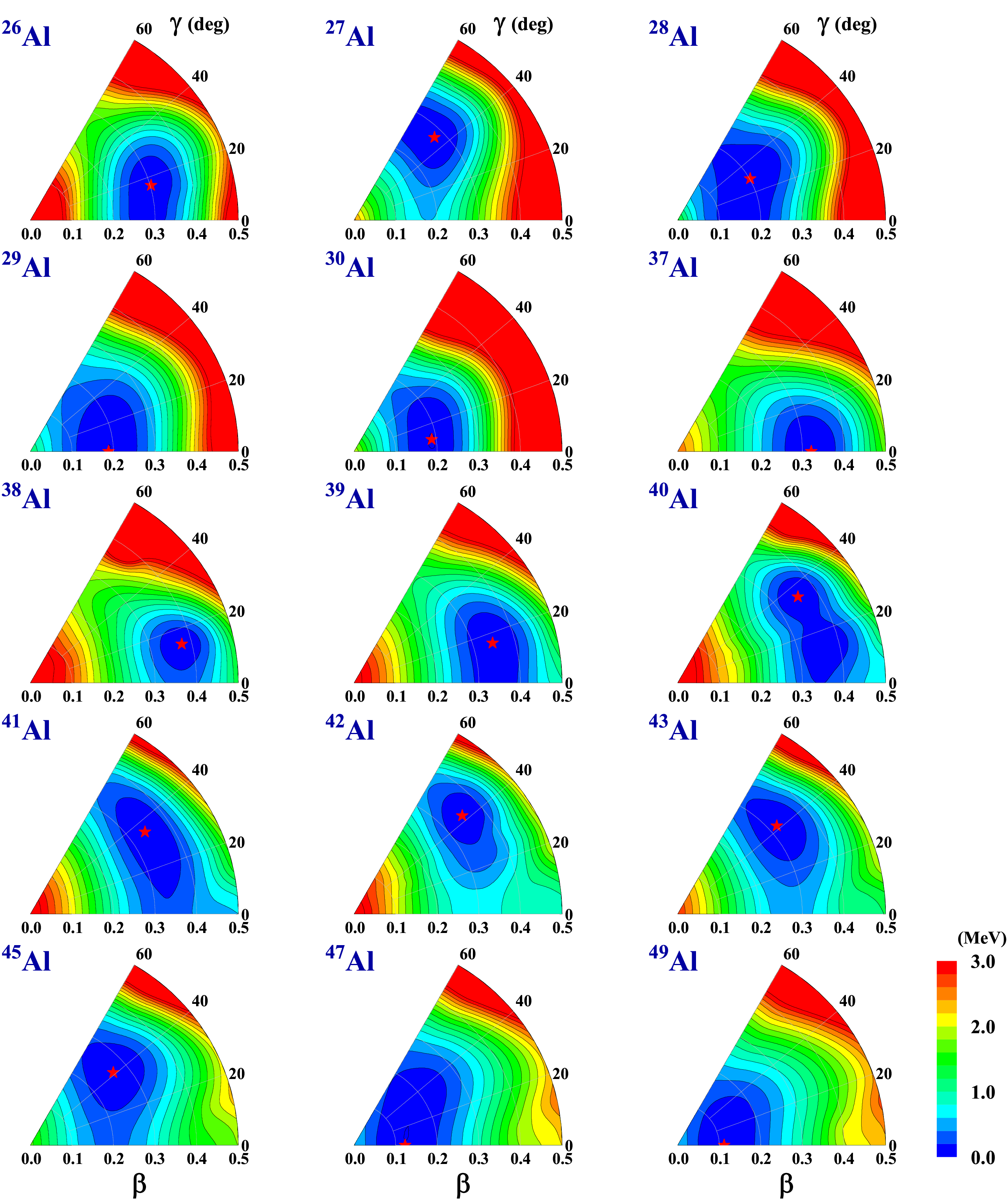}
  \caption{Potential energy surfaces for $^{26\text{--}30}$Al, $^{37\text{--}43}$Al, and $^{45,47,49}$Al from constrained TRHBc calculations with the density functional PC-PK1. All energies are normalized with respect to the energy of the absolute minimum (in MeV) indicated by the star. The energy separation between each contour line is 0.2 MeV.}
\label{fig7}
\end{figure}

\begin{figure}[htbp]
  \centering
  \includegraphics[width=0.5\textwidth]{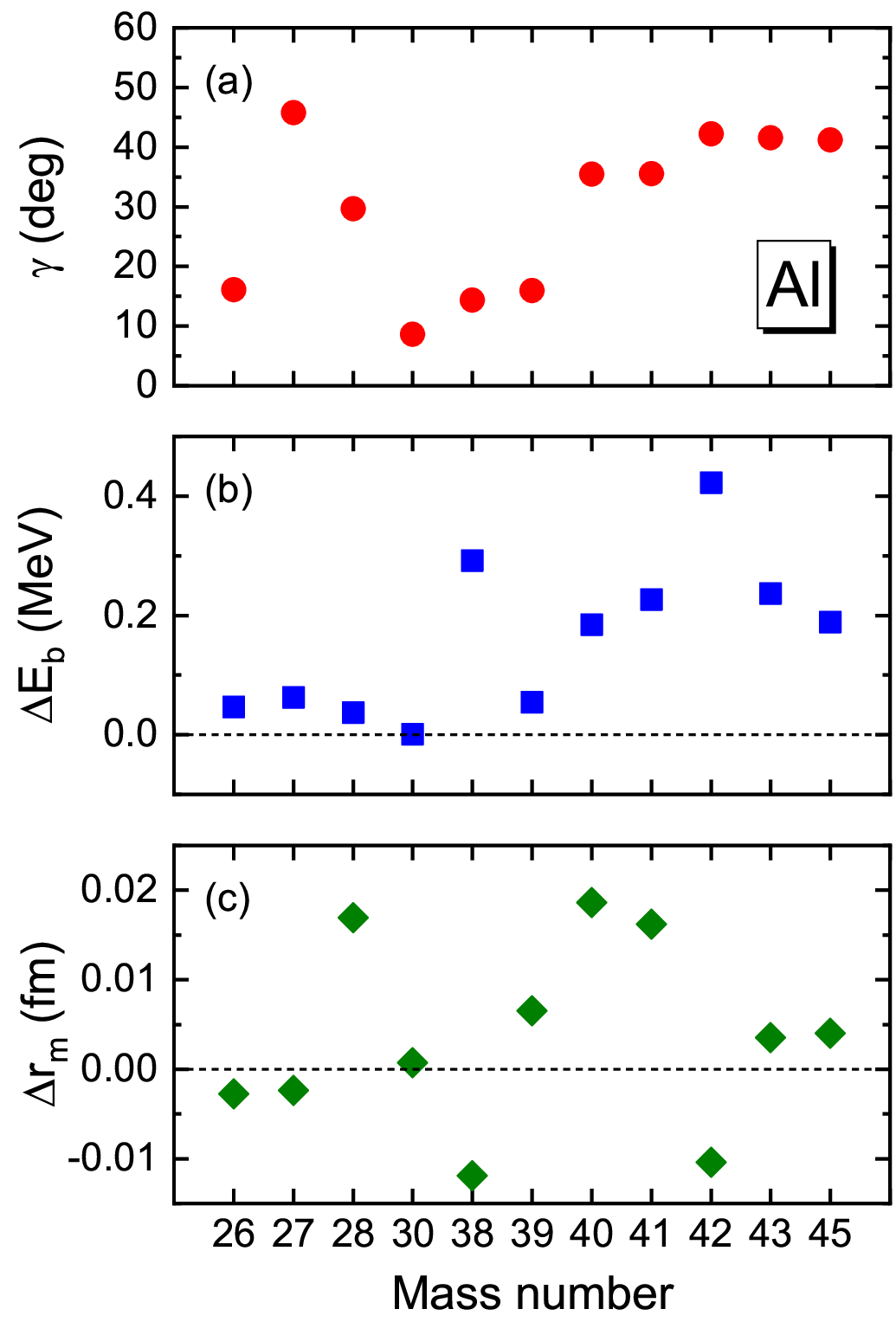}
  \caption{The triaxial deformation parameter $\gamma$ (a) and the differences between the TRHBc and DRHBc results for the binding energy (b) and for the matter rms radius (c) for the triaxially deformed nuclei $^{26\text{--}28}$Al, $^{30}$Al, $^{38\text{--}43}$Al, and $^{45}$Al.}
\label{fig8}
\end{figure}

The triaxial deformation parameters for aluminum isotopes are given in Tables \ref{tab1}--\ref{tab4}.
The predictions by the four density functionals for triaxial deformation are basically consistent.
In particular, $^{26\text{--}28}$Al, $^{30}$Al, $^{38\text{--}43}$Al, and $^{45}$Al are unambiguously predicted as triaxially deformed nuclei.
The PESs for $^{26\text{--}30}$Al, $^{37\text{--}43}$Al, and $^{45,47,49}$Al including these triaxially deformed nuclei are presented in Fig. \ref{fig7}.

To examine the triaxial deformation effects on the ground-state properties, the TRHBc and DRHBc results calculated with PC-PK1 are compared for the triaxially deformed nuclei in Fig. \ref{fig8}.
The absolute value of the DRHBc calculated deformation parameter $|\beta_2|$ is very close to the TRHBc calculated $\beta$, as demonstrated in Fig. \ref{fig7}.
Therefore, in Fig. \ref{fig8}(a), the deformation parameter $\gamma$ characterizing triaxiality is shown.
In Figs. \ref{fig8}(b) and \ref{fig8}(c), the differences between the TRHBc and DRHBc results for the binding energy $\Delta E_\mathrm{b}$ and for the matter rms radius $\Delta r_m$ are given.

In Fig. \ref{fig8}(b), all $\Delta E_\mathrm{b}$ values are positive as the inclusion of triaxial deformation increases the binding.
The value of $\Delta E_\mathrm{b}$ depends on the rigidity of the PES.
For instance, the PESs for $^{28}$Al and $^{30}$Al in Fig. \ref{fig7} are rather soft against $\gamma$, resulting in their small $\Delta E_\mathrm{b}$ shown in Fig. \ref{fig8}(b).

In Fig. \ref{fig8}(c), the triaxial deformation does not simply compress or expand the nuclear size which can also be influenced by $\Delta E_\mathrm{b}$ and the corresponding shell structure.
For $^{38}$Al and $^{42}$Al, their nuclear sizes are reduced by approximately 0.01 fm due to the large $\Delta E_\mathrm{b}$ from triaxial deformation.
For $^{28}$Al, the nuclear size is increased by approximately 0.02 fm due to the large triaxial deformation $\gamma\approx30^\circ$, with negligible $\Delta E_\mathrm{b}$.
For $^{40}$Al and $^{41}$Al, their $r_m$ are also increased by triaxial deformation effects, with moderate $\Delta E_\mathrm{b}$ values.
As shown in Ref.~\cite{Pan2025PRC} for axially deformed nuclei, the rms radii can be also impacted by the evolution of single-particle levels with respect to deformation.

\subsection{Halo scales}

In light nuclei, the halo effects can be significant, e.g., the matter radius is increased from 2.3 fm in $^{9}$Li to 3.5 fm in $^{11}$Li by adding only two neutrons.
Nevertheless, the effects of one or two halo neutrons would be less pronounced in medium-heavy and heavy nuclei.
Over the past decades, considerable efforts have been devoted to quantifying halos in medium-heavy and heavy nuclei via the examination of density profiles and nucleons in the classically forbidden region~\cite{Meng2015JPG,Zhang2019PRC,Im2000PRC,Im2000CTP}.
In Ref.~\cite{Zhang2023PRC(L2)}, a new \textit{halo scale} has been proposed,
\begin{equation}\label{HS}
 S_{\mathrm{halo}}= \frac{\Delta R^{\mathrm{exp(cal)}}}{\Delta R^{\mathrm{emp}}}= \frac{R^{\mathrm{exp(cal)}} (N + m ) - R^{\mathrm{exp(cal)}} (N)}{R^{\mathrm{emp}} (N + m ) - R^{\mathrm{emp}} (N)},
\end{equation}
where $R^{\mathrm{exp(cal)}}$ is the experimental (calculated) rms radius, $R^{\mathrm{emp}}$ is the empirical value, and $m$ is the number of halo neutrons.

\begin{figure}[htbp]
  \centering
  \includegraphics[width=0.7\textwidth]{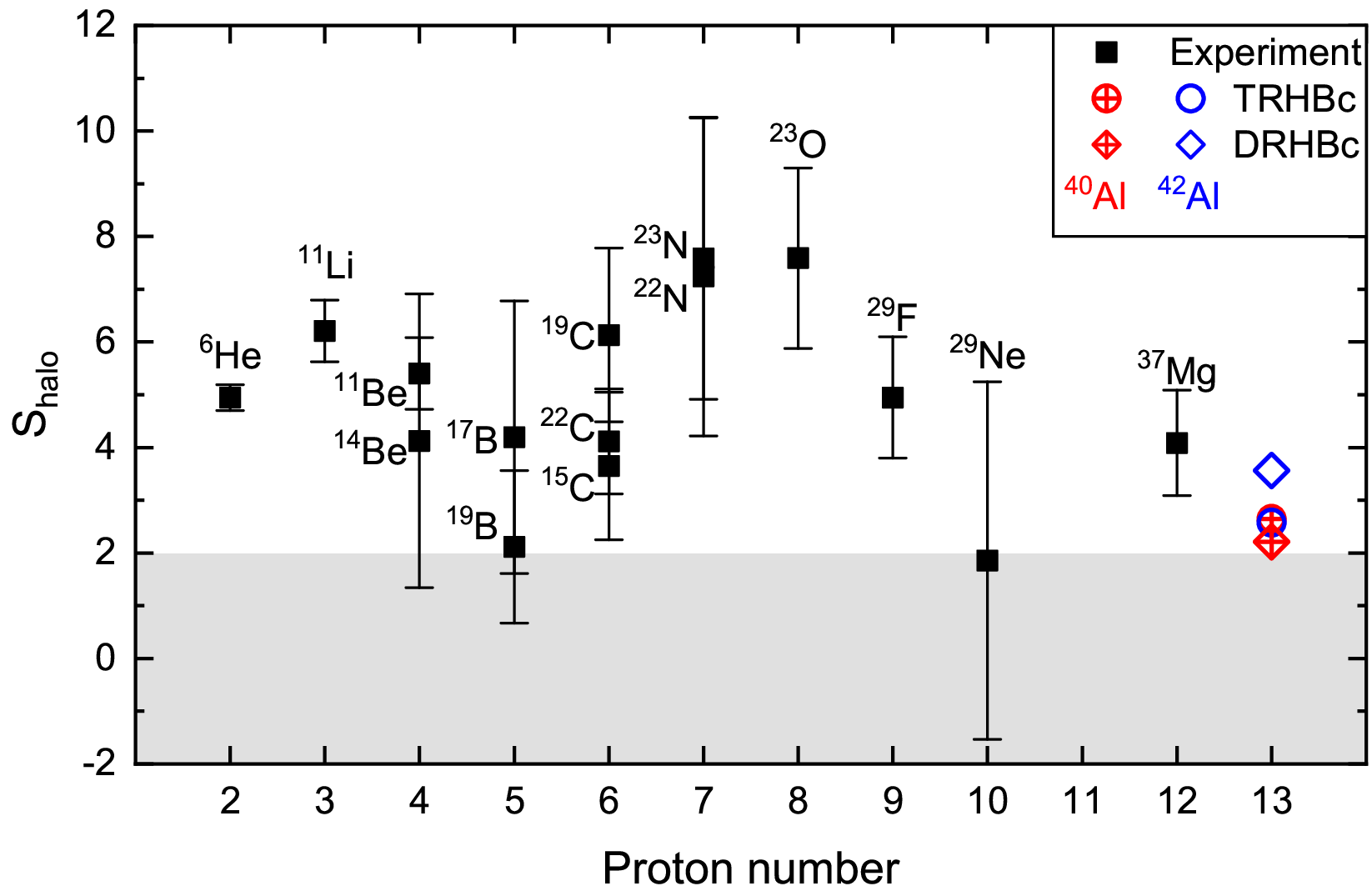}
  \caption{The halo scale $S_{\mathrm{halo}}$, for the known neutron halo nuclei and candidates extracted by using the matter radii deduced from the measured interaction cross sections~\cite{Suzuki1999NPA,Ozawa2001NPA(1),Ozawa2001NPA(2),Tanihata2013PPNP,Estrade2014PRL,Watanabe2014PRC,Kanungo2016PRL,Togano2016PLB,Bagchi2020PRL}, and for $^{40,42}$Al extracted from the TRHBc and DRHBc calculated matter radii with PC-PK1. See text for details.}
\label{fig9}
\end{figure}

Following Eq. \eqref{HS}, $S_{\mathrm{halo}}$ is computed for experimentally known neutron halo nuclei and candidates based on the matter radii deduced from the measured interaction cross sections~\cite{Suzuki1999NPA,Ozawa2001NPA(1),Ozawa2001NPA(2),Tanihata2013PPNP,Estrade2014PRL,Watanabe2014PRC,Kanungo2016PRL,Togano2016PLB,Bagchi2020PRL}. The empirical matter radius is taken as $R^{\mathrm{emp}}_m=R_0 A^{1/3}$ with $R_0$ derived from $R^{\mathrm{exp}}_m$ of the stable nucleus (referenced nucleus) in each isotopic chain.
Namely, the halo scales are extracted for $^{6}$He ($^{4}$He), $^{11}$Li ($^{6}$Li), $^{11,14}$Be ($^{9}$Be), $^{17,19}$B ($^{10}$B), $^{15,19,22}$C ($^{12}$C), $^{22,23}$N ($^{14}$N), $^{23}$O ($^{16}$O), $^{29}$F ($^{19}$F), $^{29}$Ne ($^{20}$Ne), and $^{37}$Mg ($^{24}$Mg).
The results are presented in Fig. \ref{fig9}, with the halo scales from the matter radii calculated by TRHBc and DRHBc theories for $^{40,42}$Al ($^{27}$Al) also given.

In Fig. \ref{fig9}, halo scales are generally larger than 2 for the known halo nuclei and candidates, demonstrating that the halo scale can be used to explore the emergence of halo nuclei.
Notably, the halo scales for $^{40}$Al and $^{42}$Al both exceed 2, even slightly larger than those for $^{19}$B and $^{29}$Ne.
In Ref.~\cite{Zhang2023PRC(L2)}, it was also shown that the halo scales for $^{40}$Al and $^{42}$Al are comparable to those of the predicted halo nuclei~\cite{Meng1998PRL} in zirconium isotopes.
Meanwhile, $^{40}$Al and $^{42}$Al are triaxially deformed, weakly bound nuclei, with one-neutron separation energies below 1 MeV.
Hence, it is of great interest to further explore the possible triaxial one-neutron halos in $^{40,42}$Al in a microscopic way.

\subsection{Triaxial halo nucleus $^{42}$Al}

In Ref. ~\cite{Zhang2023PRC(L2)}, the neutron-richest odd-odd aluminum isotope observed so far, $^{42}$Al, is predicted to be triaxially deformed with $\beta = 0.35$ and $\gamma=42^\circ$.
Its one-neutron separation energy is predicted to be $0.68$ MeV, in agreement with the data~\cite{AME2020(3)}, and the neutron rms radius is $3.94$ fm, remarkably larger than the empirical value.
The density distribution of the valance neutron, which extends much farther in space than the core, suggests a possible neutron halo in $^{42}$Al.
A novel phenomenon, the exchange of the intermediate and short axes between the triaxial core with $\beta =0.38$ and $\gamma = 50^\circ$ and the triaxial halo with
$\beta =0.79$ and $\gamma=-23^\circ$ is found.
By examining the single-neutron orbitals around the Fermi energy, the components responsible for the spatial extension of the halo are revealed as the approximately $46\%$ $2p$ waves, resulting in the dominant contribution of the neutron halo to the total neutron density in $^{42}$Al at large $r$.

Although the existence of a deformed one-neutron halo in $^{42}$Al is supported by the DRHBc calculations with larger $r_m$ and  $S_{\mathrm{halo}}$, the axially deformed minimum obtained in the DRHBc theory is not the ground state but a saddle point in the PES as shown in Fig. \ref{fig7}.

\subsection{Halo nucleus $^{40}$Al driven by triaxial deformation}

\begin{figure}[htbp]
    \centering
    \includegraphics[width=0.7\textwidth]{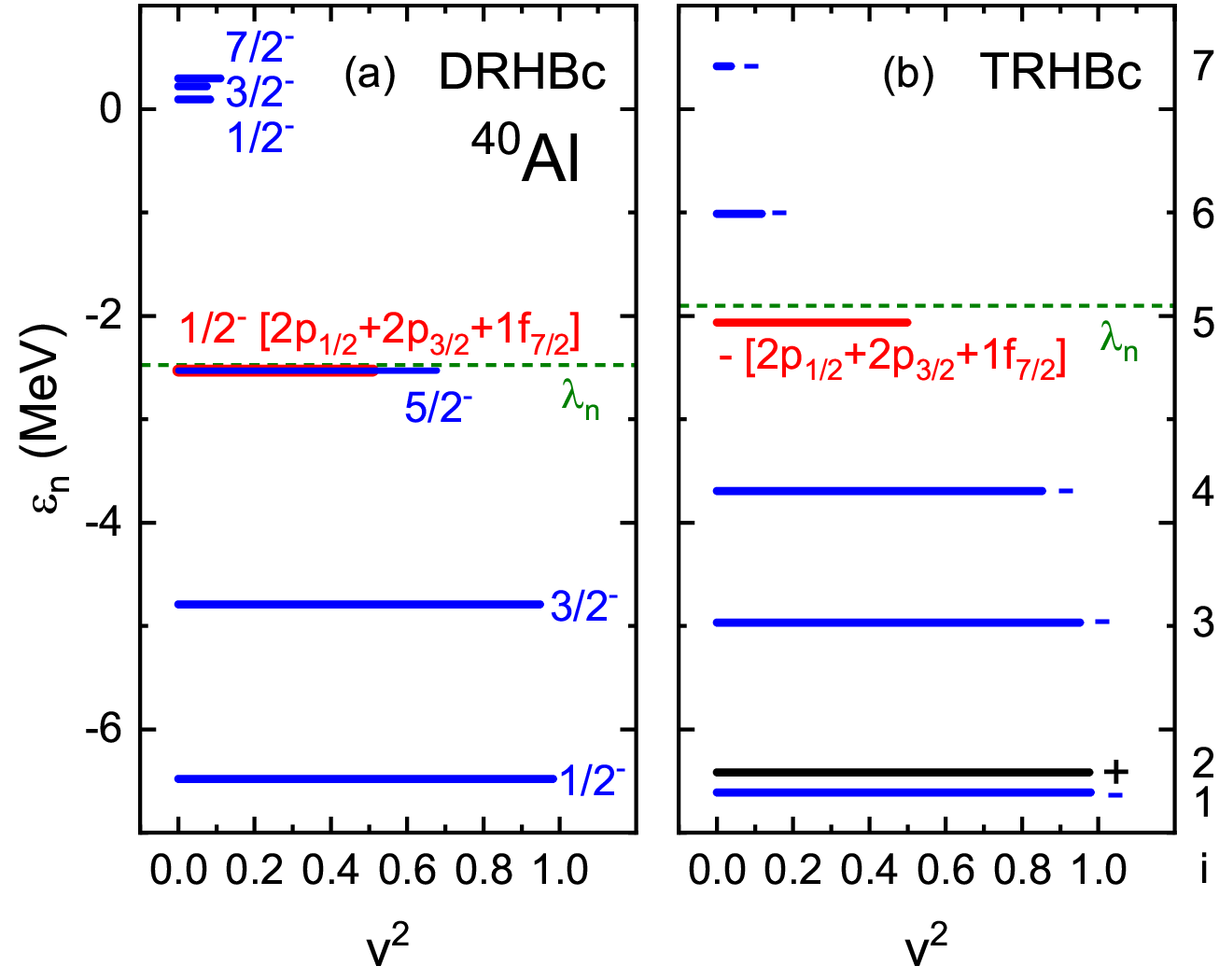}
    \caption{Single-neutron energy $\epsilon_n$ versus occupation probability $v^2$ for orbitals around the Fermi energy $\lambda_n$ (dashed line) for $^{40}$Al from the (a) DRHBc and (b) TRHBc calculations. Each orbital is labeled by quantum number(s) $\Omega^\pi$ in (a), and $\pi$ and order $i$ in (b). The orbital occupied the valence neutron is in red, with its components in the square brackets.}
    \label{fig10}
\end{figure}

As shown in Figs. \ref{fig8}(c) and \ref{fig9}, the inclusion of triaxial deformation in $^{40}$Al promotes both $r_m$ and $S_{\mathrm{halo}}$, indicating the positive role of triaxiality for the formation of a halo.
To further explore the triaxial effects, the single-neutron orbitals around the Fermi energy in the canonical basis for $^{40}$Al obtained from the DRHBc and TRHBc calculations are shown in Fig. \ref{fig10}.
The ground-state deformation parameters are $\beta_2=0.35$ in DRHBc and ($\beta = 0.36$, $\gamma=35^\circ$) in TRHBc.
In Fig. \ref{fig10}(a), each orbital is labeled by quantum numbers $\Omega^\pi$, where $\Omega$ is the third component of the angular momentum and $\pi$ is the parity.
In Fig. \ref{fig10}(b), $\Omega$ is no longer a good quantum number due to the breaking of axial symmetry, and each orbital is labeled by $\pi$ and order $i$.
In both cases, the valence neutron in $^{40}$Al occupies a weakly bound orbital with more than $60\%$ $p$-wave components.
The low centrifugal barrier for $p$ wave allows the wave function of the valence neutron to tunnel considerably into the classically forbidden region.
However, it has been recognized~\cite{Zhou2010PRC(R),Li2012PRC,Meng2015JPG,Sun2018PLB,Zhang2023PRC(L1),Zhang2023PLB,Zhang2023PRC(L2)} that a halo is well-defined only when the valence orbital decouples from other relatively deeply bound orbitals in terms of both spatial distribution and energy.
In Fig. \ref{fig10}(a),  the $1/2^-$ orbital occupied by the valence neutron is almost degenerate with an occupied $5/2^-$ orbital.
While in Fig. \ref{fig10}(b), with the inclusion of triaxial deformation, the $1/2^-$ orbital, now orbital 5, becomes more weakly bound, and the $5/2^-$ orbital, now orbital 4, becomes more deeply bound.
A sizable gap facilitating the separation between the valence neutron and the core is therefore developed.
Compared with the DRHBc case, the weaker binding of orbital 5 in the TRHBc will lead to larger $r_m$ and $S_{\mathrm{halo}}$.

\begin{figure}[htbp]
    \centering
    \includegraphics[width=0.7\textwidth]{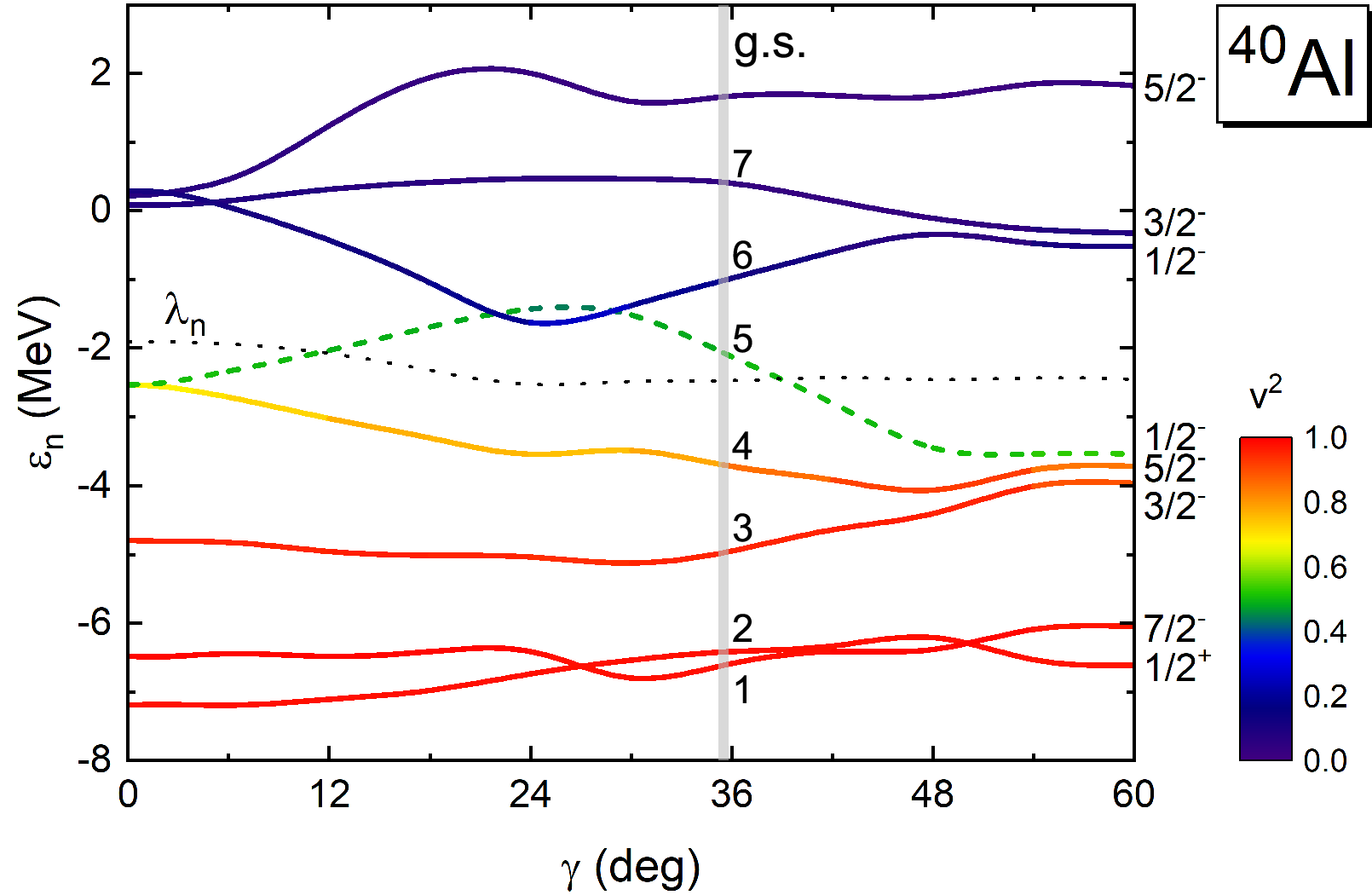}
    \caption{Single-neutron orbitals around the Fermi energy $\lambda_n$ (dotted line) of $^{40}$Al in the canonical basis as functions of the triaxial deformation parameter $\gamma$. The quantum numbers $\Omega^\pi$ in the axially deformed limit are labeled on the right. The occupation probability $v^2$ is scaled by colors. The orbital occupied by the valence neutron is shown by a dashed line. The grey vertical line corresponds to the ground state (g.s.) of $^{40}$Al, with the order $i$ for each orbital labeled.}
    \label{fig11}
\end{figure}

In Fig. \ref{fig11}, the evolution of single-neutron orbitals around the Fermi energy with respect to the deformation parameter $\gamma$ is presented.
For each $\gamma$, $\beta$ is determined by minimizing the total energy of the system.
On both the prolate and oblate sides, the halo orbital 5 and the core orbital 4 are near degenerate.
Switching on the triaxiality, this near degeneracy is broken, and the energy difference between the two orbitals grows until reaching the maximum at $\gamma\approx 27^\circ$.
For the ground state with $\gamma\approx 35^\circ$, the energy gap is around 1.6 MeV, which indicates a significant decoupling between the valence neutron and the core driven by triaxial deformation.

\begin{figure}[htbp]
    \centering
    \includegraphics[width=0.6\textwidth]{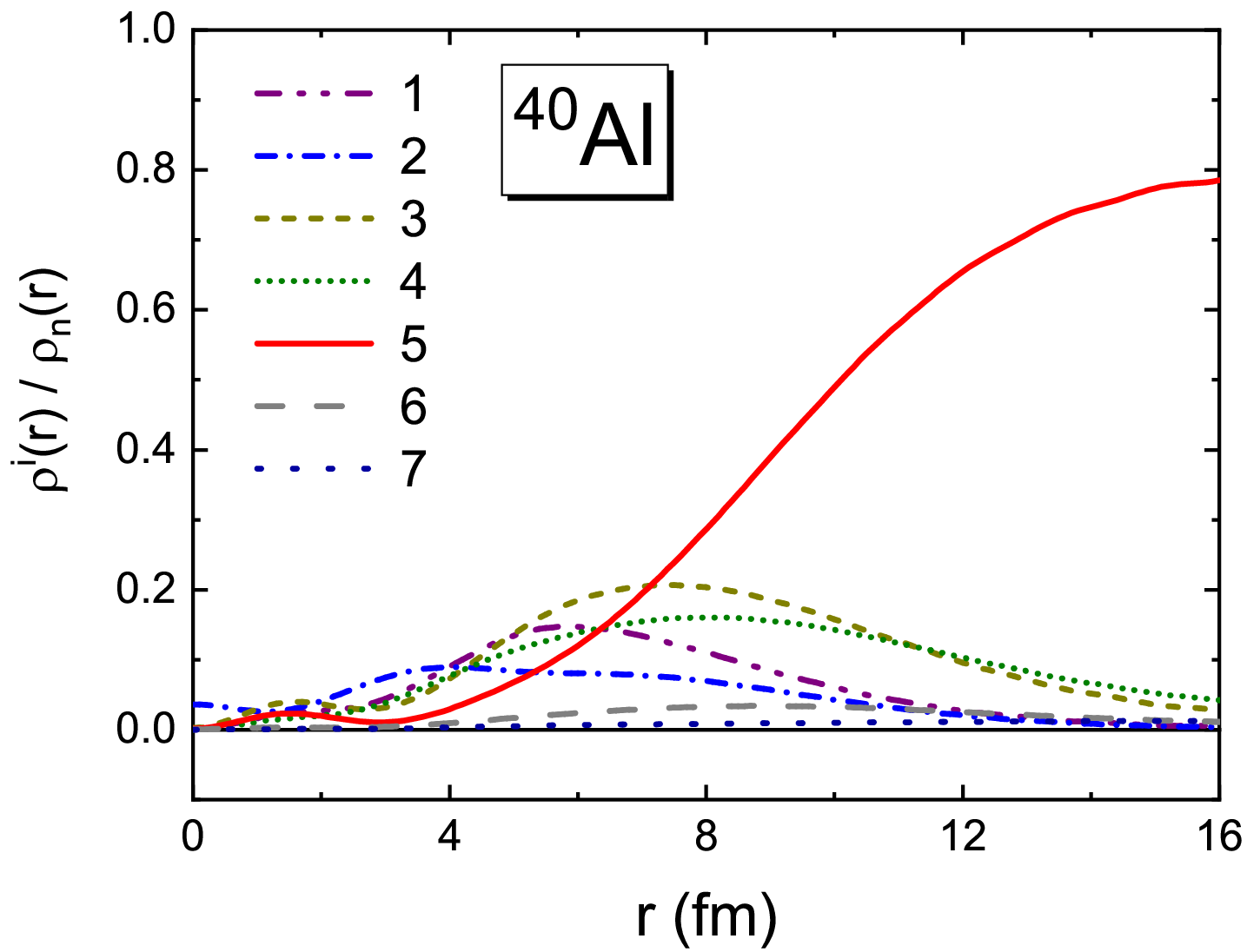}
    \caption{The contribution of each single-neutron orbital around the Fermi energy to the total neutron density as a function of $r$ for $^{40}$Al.}
    \label{fig12}
\end{figure}

In Fig. \ref{fig12}, the contributions of orbitals around the Fermi energy to the total neutron density are presented.
The contribution of orbital 5 becomes dominant beyond $r\approx 8$ fm, due to its weak binding and dominant $p$-wave components.
The contributions of orbitals 6 and 7 are quite limited due to their small occupation probabilities.
The contributions of orbitals 1--4 at large $r$ are suppressed by their deeper binding. 
Therefore, the valence neutron in $^{40}$Al dominates the total neutron density at distances far from the nuclear center, consistent with the characteristic features of nuclear halos.
The total neutron density can now be decomposed into the core and the halo, which are contributed by orbitals 1--4 and below and by orbital 5 and above, respectively.

\begin{figure}[htbp]
    \centering
    \includegraphics[width=1.0\textwidth]{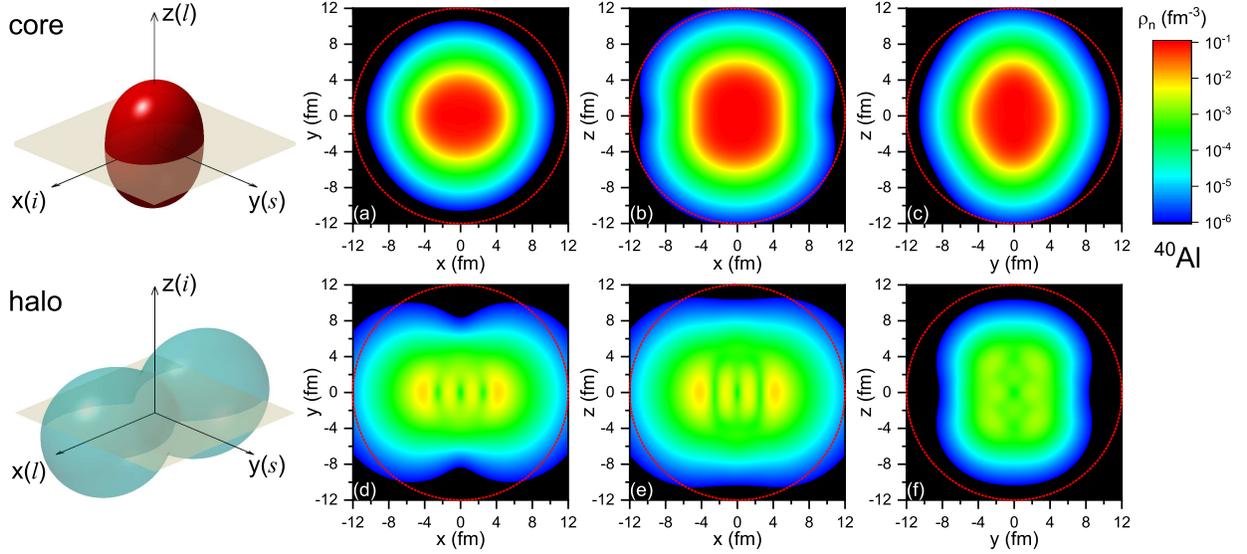}
    \caption{Neutron density distributions in $xy$, $xz$, and $yz$ planes for the core (a)--(c) and the halo (d)--(f) in the predicted triaxial halo nucleus $^{40}$Al. In each plot, a circle in dotted line is drawn to guide the eye. With the rms radius and deformation parameters $\beta$ and $\gamma$ from the densities, the corresponding schematic shapes for the core and the halo are given in the left, in which $s$, $i$, and $l$ respectively represent the short, intermediate, and long axes.}
    \label{fig13}
\end{figure}

Shown in Fig.~\ref{fig13} are the density distributions in $xy$, $xz$, and $yz$ planes for the core and the halo of $^{40}$Al.
It is remarkable that the halo density appears considerably more diffuse than the core density, particularly along the $x$ axis.
Quantitatively, the rms radii for the core and the halo are 3.73 and 5.10 fm, respectively.
The core is triaxially deformed with $\beta = 0.38$ and $\gamma=25^\circ$.
Interestingly, the deformation parameters for the halo are $\beta = 1.48$ and $\gamma=113^\circ$.
The large $\beta$ value means that the halo exhibits strong deformation with an extreme length-to-width ratio.
With the corresponding rms radius, $\beta$, and $\gamma$, schematic pictures are also given in Fig.~\ref{fig13}, where the short, intermediate, and long axes can be clearly distinguished.
Notably, an exchange of the long and intermediate axes occurs between the halo and the core.
This differs from the exchange predicted between the intermediate and short axes in $^{42}$Al~\cite{Zhang2023PRC(L2)}.
Therefore, $^{40}$Al emerges as another halo candidate featuring a unique triaxial shape decoupling, with the triaxially deformed halo enveloping the triaxially deformed core, characterized by an exchange of their long and intermediate axes.

\section{Summary}\label{summary}

In summary, a triaxial relativistic Hartree-Bogoliubov theory in continuum has been developed to incorporate triaxial deformation, pairing correlations, and continuum effects in a fully microscopic and self-consistent way, aiming for a reliable description of triaxial exotic nuclei with extreme neutron-to-proton ratios.
The TRHBc formalism is presented in detail.
The triaxial RHB equations are solved in a DWS basis, enabling a proper description for the possible large spatial extension of exotic nuclei induced by continuum effects.
The nuclear densities and potentials are expanded in spherical harmonics, allowing the inclusion of triaxial deformation degrees of freedom.

The numerical details of the TRHBc theory are presented.
The convergence for the triaxial nucleus $^{86}$Ge has been examined with respect to the box size $R_{\mathrm{box}}$, the mesh size $\Delta r$, the energy cutoff $E_{\mathrm{cut}}$, and the angular momentum cutoff $J_{\mathrm{cut}}$ for the basis as well as the spherical harmonic expansion truncation $\lambda_{\mathrm{max}}$.
No adjustment of free parameters need to be introduced in the TRHBc calculations.

The numerical implementation is benchmarked against the results from the DRHBc theory and the TRHB theory in harmonic oscillator expansion.
For $^{86}$Ge, the DRHBc calculated ground state matches the saddle point with $\gamma = 0^\circ$ on the PES from the TRHBc calculations.
The self-consistency between the ground state from unconstrained TRHBc calculations and the PES from constrained ones is also justified.
The TRHBc result is consistent with that from the TRHB calculation employing the harmonic oscillator basis.

The TRHBc theory has been applied to systematically study aluminum isotopes with density functionals PC-PK1, NL3$^*$, NLSH, and PK1.
The results exhibit reasonable agreement with available data for binding energies, one- and two-neutron separation energies, and charge radii.
The experimental proton drip-line nucleus $^{22}$Al is correctly reproduced, and the one-neutron drip line is predicted as $^{43}$Al.
The two-neutron drip line is predicted as $^{49}$Al by PC-PK1 and NL3$^*$, while as $^{47}$Al by NLSH and PK1.

The triaxially deformed nuclei $^{26\text{--}28}$Al, $^{30}$Al, $^{38\text{--}43}$Al, and $^{45}$Al are unambiguously predicted.
The triaxial deformation effects on their ground-state properties have been investigated through the comparison with the DRHBc results.
The energy gained from triaxial deformation depends on the rigidness of the PES, and the triaxial deformation does not simply compress or expand the nuclear size which can also be influenced by the corresponding binding and the shell structure.

The halo scales are computed for the known neutron halo nuclei and candidates as well as the two heaviest odd-$N$ aluminum isotopes, $^{40}$Al and $^{42}$Al, which are triaxially deformed and weakly bound with one-neutron separation energies below 1 MeV.
The halo scales for $^{40}$Al and $^{42}$Al are even larger than those for the known halo nuclei $^{19}$B and $^{29}$Ne, indicating the possible one-neutron halo structures.

Although both TRHBc and DRHBc theories predict $^{42}$Al as a deformed halo nucleus, only the TRHBc theory supports a neutron halo in $^{40}$Al.
In the TRHBc results, a sizable gap facilitating the separation between the valence neutron and the core is developed, indicating a halo driven by triaxial deformation.
A novel shape decoupling between the core and the halo has been found in $^{40}$Al, which involves not only a significant deformation change from $\beta = 0.38$ to $1.48$ but also an exchange of the long and intermediate axes with $\gamma$ from $25^\circ$ to $113^\circ$.

Extensions such as the collective Hamiltonian method, the finite amplitude method, and the angular momentum projection can be implemented in future works based on the TRHBc theory to explore the shape fluctuations as well as the vibrational and rotational excitations in triaxial exotic nuclei.
More exotic shape degrees of freedom can be included by releasing the restrictions on $\lambda$ and $\mu$ in the spherical harmonic expansion.

\begin{acknowledgments}
Helpful discussions with L. S. Geng, C. Pan, P. Ring, B. H. Sun, D. Vretenar, P. W. Zhao, S.-G. Zhou, and members of the DRHBc Mass Table Collaboration are highly appreciated. This work was partly supported by the National Natural Science Foundation of China (Grant Nos.~12305125 and  12435006), the Sichuan Science and Technology Program (Grant No. 2024NSFSC1356), the National Key Laboratory of Neutron Science and Technology (Grant No. NST202401016), the National Key R\&D Program of China (Grant No. 2024YFE0109803), and the High-performance Computing Platform of Peking University.
\end{acknowledgments}

\section*{Data Availability}

The data that support the findings of this article are openly available~\cite{data}.

\appendix

\section{Symmetry analysis}\label{symm}

In the TRHBc theory, the spatial reflection symmetry and mirror symmetries with respect to the $xy$, $xz$, and $yz$ planes are assumed.
Symmetry analysis can provide limitations on $\lambda$ and $\mu$ in the spherical harmonic expansion in Eq.~(\ref{legendre}).
Under the spatial reflection transformation,
\begin{equation}
\begin{split}
\hat P f(\bm r) &=\hat P \sum_{\lambda \mu} f_{\lambda \mu}(r) Y_{\lambda\mu}(\theta,\varphi)\\
& = \sum_{\lambda \mu} f_{\lambda \mu}(r) Y_{\lambda\mu}(\pi-\theta,\pi+\varphi)\\
& = \sum_{\lambda \mu} f_{\lambda \mu}(r) (-1)^\lambda Y_{\lambda\mu}(\theta,\varphi).
\end{split}
\end{equation}
Thus $\lambda$ is restricted to be even numbers by the spatial reflection symmetry.
This limitation also applies to the Legendre expansion in the DRHBc theory~\cite{Zhou2010PRC(R),Li2012PRC}.
Under the mirror transformation with respect to the $xy$ plane,
\begin{equation}
\begin{split}
\hat P_z V(\bm r) &=\hat P_z \sum_{\lambda \mu} f_{\lambda \mu}(r) Y_{\lambda\mu}(\theta,\varphi)\\
& = \sum_{\lambda \mu} f_{\lambda \mu}(r) Y_{\lambda\mu}(\pi-\theta,\varphi)\\
& = \sum_{\lambda \mu} f_{\lambda \mu}(r) (-1)^{\lambda+\mu} Y_{\lambda\mu}(\theta,\varphi).
\end{split}
\end{equation}
Thus $\lambda+\mu$ is restricted to be even numbers, further leading to the limitation that $\mu$ can only be even numbers.
Under the mirror transformation with respect to the $yz$ plane,
\begin{equation}
\begin{split}
\hat P_x V(\bm r) &=\hat P_x \sum_{\lambda \mu} f_{\lambda \mu}(r) Y_{\lambda\mu}(\theta,\varphi)\\
& = \sum_{\lambda \mu} f_{\lambda \mu}(r) Y_{\lambda\mu}(\theta,\pi-\varphi)\\
& = \sum_{\lambda \mu} f_{\lambda \mu}(r) Y_{\lambda-\mu}(\theta,\varphi).
\end{split}
\end{equation}
Thus the $+\mu$ component equals to the $-\mu$ one, i.e., $f_{\lambda\mu}(r) = f_{\lambda-\mu}(r)$.
The mirror symmetry with respect to the $xz$ plane also gives this limitation,
as these symmetries, satisfying $\hat P = \hat P_x \hat P_y \hat P_z$, are not fully independent.
Finally, the spherical harmonic expansion in the TRHBc theory is reduced to the simplified form in Eq.~(\ref{SHF}).

\section{Dirac Woods-Saxon basis}\label{DWS}

More explicitly, the wave function of the DWS basis in Eq.~(\ref{WFWS}) is written as
\begin{equation}
\varphi_{n\kappa m}(\bm{r} s)=\frac{1}{r}\left( \begin{matrix}
iG_{n\kappa}(r) \mathcal{Y}_{jm}^{l(1)}(\Omega,s) \\
-F_{n\kappa}(r) \mathcal{Y}_{jm}^{l(2)}(\Omega,s)
\end{matrix}\right),
\end{equation}
in which $\mathcal{Y}_{jm}^l$ is the spinor spherical harmonic,
\begin{equation}
\mathcal{Y}_{j m}^l(\Omega,s)=\sum_{m_l,m_s} \langle\frac{1}{2}m_s l m_l| j m\rangle Y_{l m_l}(\Omega)\chi_{\frac{1}{2}m_s}(s).
\end{equation}
Here, $Y_{l m_l}$ is the spherical harmonic function, $l$ is the orbital angular momentum with $m_l$ being its third component, and $\chi_{\frac{1}{2}m_s}$ is the spin wave function.
The orbital angular momenta for upper and lower components are $l(1)=j+\text{sgn}(\kappa)/2$ and $l(2)=j-\text{sgn}(\kappa)/2$, respectively.

The DWS basis is obtained by solving the Dirac equation containing spherical Woods-Saxon potentials with box boundary conditions,
\begin{equation}\label{h0}
h_0 \varphi_{n\kappa m} = \{\bm{\alpha}\cdot\bm{p}+V_{\mathrm{WS}}(r)+\beta[M+S_{\mathrm{WS}}(r)]\}\varphi_{n\kappa m} = \epsilon_{n\kappa}\varphi_{n\kappa m},
\end{equation}
where $\epsilon_{n\kappa}$ is the eigenenergy.
Using the method of separation of variables, one obtains the radial Dirac equations for radial wave functions,
\begin{equation}
\begin{aligned}
(-\frac{\partial}{\partial r}+\frac{\kappa}{r})F_{n\kappa} + [V_{\mathrm{WS}}(r) + S_{\mathrm{WS}}(r) + M] G_{n\kappa} & = \epsilon_{n\kappa} G_{n\kappa}, \\
(+\frac{\partial}{\partial r}+\frac{\kappa}{r})G_{n\kappa} + [V_{\mathrm{WS}}(r) - S_{\mathrm{WS}}(r) - M] F_{n\kappa} & = \epsilon_{n\kappa} F_{n\kappa}.
\end{aligned}
\label{rDirac}
\end{equation}
The parameterized Woods-Saxon potentials read
\begin{equation}
\begin{aligned}
V_{\mathrm{WS}}+S_{\mathrm{WS}} = \frac{V^+}{1+e^{(r-R^+)/a^+}},\\
V_{\mathrm{WS}}-S_{\mathrm{WS}} = \frac{V^-}{1+e^{(r-R^-)/a^-}},
\end{aligned}
\end{equation}
where $V^\pm$, $R^\pm$, and $a^\pm$ depict the depth, width, and diffuseness of the potential, respectively.
A usual choice of the parametrization can be found in Ref.~\cite{Koepf1991ZPA}.
For completeness, states in both the Fermi sea and the Dirac sea should be included in the DWS basis space~\cite{Zhou2003PRC}.
In some cases, the number of bases in the Dirac sea required for convergence is not small, resulting in a high computational cost~\cite{Geng2022PRC}.
In order to cure this problem for large-scale calculations, an optimized DWS basis has been recently proposed~\cite{Zhang2022PRC}.

\section{Matrix elements of the RHB Hamiltonian}\label{ME}

The Dirac Hamiltonian in Eq.~(\ref{hD}) can be written as
\begin{equation}\label{hdme}
h_D= h_0 + \sum_{\lambda\mu}[\beta S_{\lambda\mu}(r)+V_{\lambda\mu}(r)]Y_{\lambda\mu}(\Omega),
\end{equation}
where $h_0$ is the Hamiltonian for the DWS basis and $S_{\lambda\mu}$ ($V_{\lambda\mu}$) is the radial component for the spherical harmonic expansion of the scalar (vector) potential.
Here, the $\lambda=\mu=0$ component should be $[\beta (S_{00}(r)-S_{\mathrm{WS}}(r))+(V_{00}(r)-V_{\mathrm{WS}}(r))]$ but is not explicitly shown for simplicity.
Then, a matrix element of the Dirac Hamiltonian in the DWS basis is calculated by
\begin{equation}
\begin{split}
& \langle n\kappa m | h_D| n'\kappa' m'\rangle \\
= & \epsilon_{n\kappa}\delta_{nn'}\delta{\kappa\kappa'}+\langle n\kappa m | \sum_{\lambda\mu}[\beta S_{\lambda\mu}(r)+V_{\lambda\mu}(r)]Y_{\lambda\mu}(\Omega)| n'\kappa' m'\rangle \\
= & \epsilon_{n\kappa}\delta_{nn'}\delta{\kappa\kappa'} \\
& + \sum_{\sigma}\int d^3 \bm r \left\{\left(\begin{matrix}i\frac{G_{n\kappa}(r)}{r}Y_{\kappa m}^{l(1)}(\Omega, \sigma)\\-\frac{F_{n\kappa}(r)}{r}Y_{\kappa m}^{l(2)}(\Omega,\sigma)\end{matrix}\right)^\dagger\sum_{\lambda\mu}[\beta S_{\lambda\mu}(r)+V_{\lambda\mu}(r)]Y_{\lambda\mu}(\Omega)\left(\begin{matrix}i\frac{G_{n'\kappa'}(r)}{r}Y_{\kappa' m'}^{l'(1)}(\Omega, \sigma)\\-\frac{F_{n'\kappa'}(r)}{r}Y_{\kappa' m'}^{l'(2)}(\Omega,\sigma)\end{matrix}\right)\right\} \\
= & \epsilon_{n\kappa}\delta_{nn'}\delta{\kappa\kappa'} + \\
& \sum_{\lambda\mu}\langle \kappa m| Y_{\lambda\mu}|\kappa' m'\rangle \int \left\{ G_{n\kappa}(r)[V_{\lambda\mu}(r)+S_{\lambda\mu}(r)]G_{n'\kappa'}(r) + F_{n\kappa}(r)[V_{\lambda\mu}(r)-S_{\lambda\mu}(r)]F_{n'\kappa'}(r)\right\} dr,
\end{split}
\end{equation}
where the angular part $\langle \kappa m| Y_{\lambda\mu}|\kappa' m'\rangle$ can be derived with the help of the Wigner-Eckart theorem~\cite{Edmonds1957Book} and the radial integration requires numerical calculation.

For the pairing channel, since the pairing interaction in Eq.~(\ref{pair}) projects onto the spin $S=0$ component, one can find~\cite{Li2012PRC}
\begin{equation}
\sum_s (-)^{\frac{1}{2}-s}\varphi_{n\kappa m}(s)\overline\varphi_{n'\kappa'm'}(-s) = \sum_s \varphi_{n\kappa m}(s)\overline\varphi_{n'\kappa'm'}^*(s) = \sum_{\lambda\mu} \frac{R_{n\kappa}R_{n'\kappa'}}{r^2}\langle\kappa m|Y^*_{\lambda\mu}|\kappa'm'\rangle Y_{\lambda\mu}.
\end{equation}
Therefore, the matrix elements of the pairing potential can be calculated by
\begin{equation}
\langle n\kappa m |\Delta^{++}|\overline{n'\kappa'm'}\rangle = \sum_{\lambda\mu} \langle\kappa m|Y_{\lambda\mu}|\kappa'm'\rangle \int dr G_{n\kappa}\Delta_{\lambda\mu}(r)G_{n'\kappa'},
\end{equation}
and
\begin{equation}
\langle n\kappa m |\Delta^{--}|\overline{n'\kappa'm'}\rangle = \sum_{\lambda\mu} \langle\kappa m|Y_{\lambda\mu}|\kappa'm'\rangle \int dr F_{n\kappa}\Delta_{\lambda\mu}(r)F_{n'\kappa'},
\end{equation}
where $\Delta_{\lambda\mu}(r)$ is the radial component for the spherical harmonic expansion of the pairing potential and will be given below.
Here, the pairing field connecting upper and lower components have been assumed to be zero, i.e., $\Delta^{+-}(\bm r) = \kappa^{+-}(\bm r) =0$.
As demonstrated in Ref.~\cite{Serra2002PRC}, $\Delta^{+-}$ is an order of magnitude smaller than $\Delta^{++}$ and, thus, can be neglected as a very good approximation.

The radial components of the pairing tensor are calculated as
 \begin{equation}
\kappa_{\lambda\mu}^{++} = \frac{1}{r^2} \sum_{n\kappa m,n'\kappa'm'} G_{n\kappa} \kappa_{(n\kappa m),(n'\kappa'm')} G_{n'\kappa'} \langle\kappa m|Y_{\lambda\mu}^*|\kappa'm'\rangle,
\end{equation}
 \begin{equation}
\kappa_{\lambda\mu}^{++} = \frac{1}{r^2} \sum_{n\kappa m,n'\kappa'm'} F_{n\kappa} \kappa_{(n\kappa m),(n'\kappa'm')} F_{n'\kappa'} \langle\kappa m|Y_{\lambda\mu}^*|\kappa'm'\rangle,
\end{equation}
where
\begin{equation}
\kappa_{(n\kappa m),(n'\kappa'm')} = \sum_{k>0} v_{k,(n\kappa m)}^* u_{k,(n'\kappa'm')},
\end{equation}
is the matrix element of the pairing tensor $\kappa = V^* U^\mathrm{T}$ in the DWS basis.

Finally, the pairing potential in Eq.~(\ref{Delta}) has the local form
\begin{equation}
\Delta^{++(--)}(\bm r) = V_0 [1-\frac{\rho(\bm r)}{\rho_{\mathrm{sat}}}] \kappa^{++(--)}(\bm r),
\end{equation}
whose radial component for the spherical harmonic expansion is calculated as
\begin{equation}
\Delta^{++(--)}_{\lambda\mu}(r) = V_0 \int d\Omega Y_{\lambda\mu}^* [1-\frac{1}{\rho_{\mathrm{sat}}}\sum_{\lambda_1\mu_1}\rho_{\lambda_1\mu_1}(r) Y_{\lambda_1\mu_1}] \sum_{\lambda_2\mu_2}\kappa^{++(--)}_{\lambda_2\mu_2}(r) Y_{\lambda_2\mu_2},
\end{equation}
which can be reduced by using
\begin{equation}
\int d\Omega Y_{\lambda\mu}^* Y_{\lambda_1\mu_1}Y_{\lambda_2\mu_2} = \sqrt{\frac{(2\lambda_1+1)(2\lambda_2+1)}{4\pi(2\lambda+1)}} \langle\lambda_1\mu_1\lambda_2\mu_2|\lambda\mu\rangle \langle\lambda_1 0\lambda_2 0 |\lambda 0\rangle.
\end{equation}

\section{Tabulation of ground-state properties for bound aluminum isotopes}\label{tables}

\begin{table}[H]
  \centering
  \caption{Ground-state properties including the binding energy $E_\mathrm{b}^\mathrm{cal}$, neutron rms radius $R_n$, proton rms radius $R_p$, matter rms radius $R_m$, charge radius $R_\mathrm{ch}^\mathrm{cal}$, and deformation parameters $\beta$ and $\gamma$ for the bound aluminum isotopes calculated by the TRHBc theory with the density functional PC-PK1. Available experimental values for the binding energy $E_\mathrm{b}^\mathrm{exp}$~\cite{AME2020(3)} and the charge radius $R_\mathrm{ch}^\mathrm{exp}$~\cite{Heylen2021PRC} are listed for comparison, with the rms deviations from the data given at the bottom.}
  {\renewcommand{\arraystretch}{0.8}
  \begin{tabular}{ccccccccccc}
    \hline\hline
    $A$ & $N$ & $E_\mathrm{b}^\mathrm{cal}$ & $E_\mathrm{b}^\mathrm{exp}$ & $R_n$ & $R_p$ & $R_m$ & $R_\mathrm{ch}^\mathrm{cal}$  & $R_\mathrm{ch}^\mathrm{exp}$ & $\beta$ & $\gamma$ \\
     &  &  (MeV) &  (MeV) & (fm)& (fm)& (fm)& (fm) & (fm) & & (deg) \\
    \hline
22	&	9	&	149.92 	&		&	2.74 	&	3.08 	&	2.94 	&	3.18 	&		&	0.29 	&	0.00 	\\
23	&	10	&	167.72 	&	168.72 	&	2.80 	&	3.04 	&	2.94 	&	3.14 	&		&	0.34 	&	0.00 	\\
24	&	11	&	183.36 	&	183.59 	&	2.85 	&	3.00 	&	2.93 	&	3.11 	&		&	0.39 	&	0.00 	\\
25	&	12	&	201.56 	&	200.53 	&	2.81 	&	2.89 	&	2.85 	&	3.00 	&		&	0.39 	&	0.00 	\\
26	&	13	&	212.20 	&	211.89 	&	2.85 	&	2.88 	&	2.87 	&	2.99 	&		&	0.30 	&	16.12 	\\
27	&	14	&	224.14 	&	224.95 	&	2.95 	&	2.93 	&	2.94 	&	3.04 	&	3.06 	&	0.28 	&	45.77 	\\
28	&	15	&	231.86 	&	232.68 	&	3.02 	&	2.93 	&	2.98 	&	3.03 	&	3.06 	&	0.20 	&	29.70 	\\
29	&	16	&	240.65 	&	242.11 	&	3.09 	&	2.95 	&	3.03 	&	3.06 	&	3.08 	&	0.19 	&	0.00 	\\
30	&	17	&	247.37 	&	247.83 	&	3.17 	&	2.98 	&	3.09 	&	3.08 	&	3.09 	&	0.19 	&	8.67 	\\
31	&	18	&	254.89 	&	254.99 	&	3.23 	&	3.00 	&	3.14 	&	3.10 	&	3.11 	&	0.16 	&	0.00 	\\
32	&	19	&	260.54 	&	259.21 	&	3.29 	&	3.02 	&	3.19 	&	3.12 	&	3.08 	&	0.11 	&	0.00 	\\
33	&	20	&	267.39 	&	264.68 	&	3.34 	&	3.04 	&	3.23 	&	3.15 	&		&	0.06 	&	0.00 	\\
34	&	21	&	270.09 	&	267.25 	&	3.42 	&	3.07 	&	3.29 	&	3.17 	&		&	0.17 	&	0.00 	\\
35	&	22	&	275.07 	&	272.55 	&	3.48 	&	3.09 	&	3.34 	&	3.19 	&		&	0.21 	&	0.00 	\\
36	&	23	&	277.80 	&	274.45 	&	3.56 	&	3.12 	&	3.40 	&	3.22 	&		&	0.32 	&	0.00 	\\
37	&	24	&	282.01 	&	278.66 	&	3.61 	&	3.13 	&	3.45 	&	3.23 	&		&	0.32 	&	0.00 	\\
38	&	25	&	283.88 	&		&	3.68 	&	3.16 	&	3.51 	&	3.26 	&		&	0.38 	&	14.35 	\\
39	&	26	&	287.21 	&		&	3.73 	&	3.17 	&	3.55 	&	3.27 	&		&	0.35 	&	16.00 	\\
40	&	27	&	288.17 	&		&	3.82 	&	3.19 	&	3.63 	&	3.29 	&		&	0.36 	&	35.49 	\\
41	&	28	&	291.18 	&		&	3.85 	&	3.20 	&	3.66 	&	3.30 	&		&	0.34 	&	35.53 	\\
42	&	29	&	291.86 	&		&	3.94 	&	3.21 	&	3.73 	&	3.31 	&		&	0.35 	&	42.22 	\\
43	&	30	&	294.21 	&		&	3.97 	&	3.22 	&	3.76 	&	3.32 	&		&	0.32 	&	41.60 	\\
45	&	32	&	296.19 	&		&	4.07 	&	3.23 	&	3.84 	&	3.33 	&		&	0.27 	&	41.20 	\\
47	&	34	&	297.50 	&		&	4.16 	&	3.21 	&	3.92 	&	3.31 	&		&	0.12 	&	0.00 	\\
49	&	36	&	298.06 	&		&	4.25 	&	3.25 	&	4.01 	&	3.35 	&		&	0.11 	&	0.00 	\\
$\sigma$	&	 &	1.86 	&		&	 	&	 	&	 	&	0.03 	&		&	 	&	 	\\
    \hline\hline
  \end{tabular}}
  \label{tab1}
\end{table}

\begin{table}[H]
  \centering
  \caption{Same as Table \ref{tab1} but with the density functional NL3$^*$.}
  {\renewcommand{\arraystretch}{0.85}
  \begin{tabular}{ccccccccccc}
    \hline\hline
    $A$ & $N$ & $E_\mathrm{b}^\mathrm{cal}$ & $E_\mathrm{b}^\mathrm{exp}$ & $R_n$ & $R_p$ & $R_m$ & $R_\mathrm{ch}^\mathrm{cal}$  & $R_\mathrm{ch}^\mathrm{exp}$ & $\beta$ & $\gamma$ \\
     &  &  (MeV) &  (MeV) & (fm)& (fm)& (fm)& (fm) & (fm) & & (deg) \\
    \hline
    22	&	9	&	149.30 	&		&	2.70 	&	3.07 	&	2.92 	&	3.17 	&		&	0.29 	&	0.00 	\\
    23	&	10	&	166.53 	&	168.72 	&	2.78 	&	3.04 	&	2.93 	&	3.14 	&		&	0.36 	&	0.00 	\\
    24	&	11	&	181.14 	&	183.59 	&	2.85 	&	3.01 	&	2.94 	&	3.11 	&		&	0.39 	&	0.00 	\\
    25	&	12	&	196.44 	&	200.53 	&	2.89 	&	2.98 	&	2.93 	&	3.08 	&		&	0.39 	&	0.00 	\\
    26	&	13	&	207.75 	&	211.89 	&	2.93 	&	2.97 	&	2.95 	&	3.07 	&		&	0.33 	&	25.44 	\\
    27	&	14	&	220.55 	&	224.95 	&	2.98 	&	2.96 	&	2.97 	&	3.07 	&	3.06 	&	0.32 	&	47.18 	\\
    28	&	15	&	228.70 	&	232.68 	&	3.04 	&	2.95 	&	3.00 	&	3.06 	&	3.06 	&	0.27 	&	30.66 	\\
    29	&	16	&	238.07 	&	242.11 	&	3.10 	&	2.96 	&	3.03 	&	3.06 	&	3.08 	&	0.23 	&	0.00 	\\
    30	&	17	&	245.06 	&	247.83 	&	3.17 	&	2.97 	&	3.09 	&	3.08 	&	3.09 	&	0.22 	&	18.97 	\\
    31	&	18	&	253.07 	&	254.99 	&	3.23 	&	2.98 	&	3.13 	&	3.09 	&	3.11 	&	0.16 	&	0.02 	\\
    32	&	19	&	258.90 	&	259.21 	&	3.29 	&	3.00 	&	3.17 	&	3.10 	&	3.08 	&	0.11 	&	0.00 	\\
    33	&	20	&	266.06 	&	264.68 	&	3.33 	&	3.02 	&	3.21 	&	3.12 	&		&	0.06 	&	0.00 	\\
    34	&	21	&	268.64 	&	267.25 	&	3.41 	&	3.04 	&	3.27 	&	3.14 	&		&	0.17 	&	0.00 	\\
    35	&	22	&	273.70 	&	272.55 	&	3.47 	&	3.06 	&	3.32 	&	3.16 	&		&	0.21 	&	0.00 	\\
    36	&	23	&	276.12 	&	274.45 	&	3.55 	&	3.09 	&	3.39 	&	3.19 	&		&	0.32 	&	0.00 	\\
    37	&	24	&	280.42 	&	278.66 	&	3.60 	&	3.10 	&	3.43 	&	3.20 	&		&	0.32 	&	0.00 	\\
    38	&	25	&	281.97 	&		&	3.68 	&	3.13 	&	3.50 	&	3.23 	&		&	0.38 	&	15.32 	\\
    39	&	26	&	285.47 	&		&	3.74 	&	3.14 	&	3.55 	&	3.24 	&		&	0.35 	&	18.73 	\\
    40	&	27	&	286.23 	&		&	3.83 	&	3.16 	&	3.63 	&	3.26 	&		&	0.36 	&	36.95 	\\
    41	&	28	&	289.28 	&		&	3.87 	&	3.17 	&	3.66 	&	3.27 	&		&	0.35 	&	37.18 	\\
    42	&	29	&	289.62 	&		&	3.96 	&	3.18 	&	3.74 	&	3.28 	&		&	0.36 	&	42.33 	\\
    43	&	30	&	292.06 	&		&	3.99 	&	3.19 	&	3.76 	&	3.29 	&		&	0.33 	&	42.03 	\\
    45	&	32	&	293.77 	&		&	4.10 	&	3.20 	&	3.86 	&	3.30 	&		&	0.28 	&	41.44 	\\
    47	&	34	&	294.71 	&		&	4.19 	&	3.18 	&	3.94 	&	3.28 	&		&	0.13 	&	0.00 	\\
    49	&	36	&	295.10 	&		&	4.30 	&	3.21 	&	4.04 	&	3.31 	&		&	0.12 	&	0.02 	\\
    $\sigma$	&	 &	2.81 	&		&	 	&	 	&	 	&	0.02 	&		&	 	&	 	\\
    \hline\hline
  \end{tabular}}
  \label{tab2}
\end{table}

\begin{table}[H]
  \centering
  \caption{Same as Table \ref{tab1} but with the density functional NLSH.}
  {\renewcommand{\arraystretch}{0.85}
  \begin{tabular}{ccccccccccc}
    \hline\hline
    $A$ & $N$ & $E_\mathrm{b}^\mathrm{cal}$ & $E_\mathrm{b}^\mathrm{exp}$ & $R_n$ & $R_p$ & $R_m$ & $R_\mathrm{ch}^\mathrm{cal}$  & $R_\mathrm{ch}^\mathrm{exp}$ & $\beta$ & $\gamma$ \\
     &  &  (MeV) &  (MeV) & (fm)& (fm)& (fm)& (fm) & (fm) & & (deg) \\
    \hline
22	&	9	&	149.84 	&		&	2.67 	&	3.01 	&	2.88 	&	3.11 	&		&	0.27 	&	0.00 	\\
23	&	10	&	167.23 	&	168.72 	&	2.76 	&	3.00 	&	2.90 	&	3.10 	&		&	0.36 	&	0.00 	\\
24	&	11	&	182.06 	&	183.59 	&	2.83 	&	2.98 	&	2.91 	&	3.09 	&		&	0.38 	&	0.00 	\\
25	&	12	&	197.48 	&	200.53 	&	2.88 	&	2.97 	&	2.92 	&	3.07 	&		&	0.38 	&	0.00 	\\
26	&	13	&	208.84 	&	211.89 	&	2.92 	&	2.95 	&	2.93 	&	3.06 	&		&	0.32 	&	22.21 	\\
27	&	14	&	221.65 	&	224.95 	&	2.97 	&	2.95 	&	2.96 	&	3.06 	&	3.06 	&	0.30 	&	47.31 	\\
28	&	15	&	229.67 	&	232.68 	&	3.01 	&	2.94 	&	2.98 	&	3.04 	&	3.06 	&	0.23 	&	14.32 	\\
29	&	16	&	238.96 	&	242.11 	&	3.08 	&	2.96 	&	3.03 	&	3.07 	&	3.08 	&	0.27 	&	0.00 	\\
30	&	17	&	245.47 	&	247.83 	&	3.16 	&	2.97 	&	3.08 	&	3.08 	&	3.09 	&	0.23 	&	22.81 	\\
31	&	18	&	253.16 	&	254.99 	&	3.21 	&	2.98 	&	3.12 	&	3.09 	&	3.11 	&	0.18 	&	27.84 	\\
32	&	19	&	258.87 	&	259.21 	&	3.26 	&	2.99 	&	3.15 	&	3.10 	&	3.08 	&	0.12 	&	29.55 	\\
33	&	20	&	266.14 	&	264.68 	&	3.30 	&	3.00 	&	3.19 	&	3.11 	&		&	0.05 	&	0.00 	\\
34	&	21	&	268.72 	&	267.25 	&	3.37 	&	3.02 	&	3.24 	&	3.13 	&		&	0.15 	&	0.00 	\\
35	&	22	&	273.66 	&	272.55 	&	3.42 	&	3.04 	&	3.29 	&	3.14 	&		&	0.20 	&	0.00 	\\
36	&	23	&	276.33 	&	274.45 	&	3.50 	&	3.07 	&	3.35 	&	3.17 	&		&	0.30 	&	0.00 	\\
37	&	24	&	280.52 	&	278.66 	&	3.55 	&	3.09 	&	3.39 	&	3.19 	&		&	0.32 	&	0.00 	\\
38	&	25	&	282.26 	&		&	3.61 	&	3.11 	&	3.45 	&	3.21 	&		&	0.35 	&	13.52 	\\
39	&	26	&	285.31 	&		&	3.67 	&	3.12 	&	3.50 	&	3.22 	&		&	0.34 	&	18.13 	\\
40	&	27	&	286.07 	&		&	3.75 	&	3.15 	&	3.57 	&	3.25 	&		&	0.35 	&	37.98 	\\
41	&	28	&	288.58 	&		&	3.80 	&	3.15 	&	3.61 	&	3.25 	&		&	0.34 	&	40.81 	\\
42	&	29	&	288.94 	&		&	3.90 	&	3.16 	&	3.69 	&	3.26 	&		&	0.35 	&	44.30 	\\
43	&	30	&	290.86 	&		&	3.92 	&	3.17 	&	3.71 	&	3.27 	&		&	0.32 	&	42.49 	\\
45	&	32	&	292.10 	&		&	4.04 	&	3.17 	&	3.81 	&	3.27 	&		&	0.27 	&	41.83 	\\
47	&	34	&	292.74 	&		&	4.15 	&	3.15 	&	3.90 	&	3.25 	&		&	0.11 	&	0.00 	\\
$\sigma$	&	 &	2.23 	&		&	 	&	 	&	 	&	0.02 	&		&	 	&	 	\\
    \hline\hline
  \end{tabular}}
  \label{tab3}
\end{table}

\begin{table}[H]
  \centering
  \caption{Same as Table \ref{tab1} but with the density functional PK1.}
  {\renewcommand{\arraystretch}{0.85}
  \begin{tabular}{ccccccccccc}
    \hline\hline
    $A$ & $N$ & $E_\mathrm{b}^\mathrm{cal}$ & $E_\mathrm{b}^\mathrm{exp}$ & $R_n$ & $R_p$ & $R_m$ & $R_\mathrm{ch}^\mathrm{cal}$  & $R_\mathrm{ch}^\mathrm{exp}$ & $\beta$ & $\gamma$ \\
     &  &  (MeV) &  (MeV) & (fm)& (fm)& (fm)& (fm) & (fm) & & (deg) \\
    \hline
22	&	9	&	149.31 	&		&	2.67 	&	3.03 	&	2.89 	&	3.13 	&		&	0.28 	&	0.00 	\\
23	&	10	&	166.94 	&	168.72 	&	2.76 	&	3.01 	&	2.91 	&	3.12 	&		&	0.37 	&	0.00 	\\
24	&	11	&	182.10 	&	183.59 	&	2.83 	&	2.99 	&	2.92 	&	3.09 	&		&	0.39 	&	0.00 	\\
25	&	12	&	197.66 	&	200.53 	&	2.89 	&	2.98 	&	2.94 	&	3.09 	&		&	0.41 	&	10.86 	\\
26	&	13	&	209.11 	&	211.89 	&	2.93 	&	2.97 	&	2.95 	&	3.07 	&		&	0.35 	&	27.22 	\\
27	&	14	&	221.99 	&	224.95 	&	2.98 	&	2.96 	&	2.97 	&	3.07 	&	3.06 	&	0.33 	&	48.30 	\\
28	&	15	&	229.91 	&	232.68 	&	3.04 	&	2.96 	&	3.00 	&	3.06 	&	3.06 	&	0.29 	&	31.53 	\\
29	&	16	&	239.20 	&	242.11 	&	3.08 	&	2.96 	&	3.03 	&	3.06 	&	3.08 	&	0.26 	&	0.00 	\\
30	&	17	&	245.87 	&	247.83 	&	3.16 	&	2.97 	&	3.08 	&	3.08 	&	3.09 	&	0.24 	&	24.16 	\\
31	&	18	&	253.42 	&	254.99 	&	3.21 	&	2.98 	&	3.12 	&	3.09 	&	3.11 	&	0.18 	&	29.42 	\\
32	&	19	&	259.31 	&	259.21 	&	3.26 	&	2.99 	&	3.15 	&	3.09 	&	3.08 	&	0.12 	&	24.73 	\\
33	&	20	&	266.54 	&	264.68 	&	3.31 	&	3.00 	&	3.19 	&	3.10 	&		&	0.06 	&	0.00 	\\
34	&	21	&	268.96 	&	267.25 	&	3.38 	&	3.02 	&	3.25 	&	3.12 	&		&	0.16 	&	0.00 	\\
35	&	22	&	273.44 	&	272.55 	&	3.44 	&	3.04 	&	3.30 	&	3.14 	&		&	0.21 	&	0.00 	\\
36	&	23	&	276.22 	&	274.45 	&	3.52 	&	3.07 	&	3.36 	&	3.17 	&		&	0.31 	&	0.00 	\\
37	&	24	&	279.98 	&	278.66 	&	3.57 	&	3.09 	&	3.41 	&	3.19 	&		&	0.32 	&	0.00 	\\
38	&	25	&	281.81 	&		&	3.65 	&	3.12 	&	3.47 	&	3.22 	&		&	0.37 	&	15.55 	\\
39	&	26	&	284.56 	&		&	3.70 	&	3.13 	&	3.52 	&	3.23 	&		&	0.36 	&	20.08 	\\
40	&	27	&	285.51 	&		&	3.79 	&	3.15 	&	3.60 	&	3.25 	&		&	0.37 	&	38.68 	\\
41	&	28	&	287.77 	&		&	3.84 	&	3.16 	&	3.64 	&	3.26 	&		&	0.35 	&	40.11 	\\
42	&	29	&	288.20 	&		&	3.94 	&	3.17 	&	3.72 	&	3.27 	&		&	0.37 	&	43.70 	\\
43	&	30	&	289.91 	&		&	3.96 	&	3.18 	&	3.74 	&	3.28 	&		&	0.34 	&	43.19 	\\
45	&	32	&	291.02 	&		&	4.08 	&	3.19 	&	3.84 	&	3.29 	&		&	0.29 	&	42.50 	\\
47	&	34	&	291.41 	&		&	4.19 	&	3.19 	&	3.94 	&	3.29 	&		&	0.20 	&	36.42 	\\
$\sigma$	&	 &	2.08 	&		&	 	&	 	&	 	&	0.01 	&		&	 	&	 	\\
    \hline\hline
  \end{tabular}}
  \label{tab4}
\end{table}


%

\end{document}